\documentclass[aps,prd,twocolumn,nofootinbib,superscriptaddress]{revtex4-2}

\usepackage{graphicx}
\usepackage{dcolumn}
\usepackage{bm}
\usepackage[normalem]{ulem}

\usepackage{xcolor}
\usepackage{booktabs}

\newcommand\ege{{Department of Astronomy and Space Sciences, Faculty of Science, Ege University, 35100, {\.I}zmir, Türkiye}}
\newcommand\egeinst{{Department of Astronomy and Space Sciences, Graduate School of Natural and Applied Sciences, Ege University, 35100, {\.I}zmir, Türkiye}}
\newcommand\egravity{Ege Gravitational Astrophysics Research Group (eGRAVITY), Ege University, 35100, {\.I}zmir, Türkiye}
\newcommand\unitn{Dipartimento di Fisica, Università di Trento, via Sommarive 14, 38123 Trento, Italy}
\newcommand\infn{INFN-TIFPA, Trento Institute for Fundamental Physics and Applications, via Sommarive 14, 38123 Trento, Italy}
\newcommand\ioa{Institute of Astronomy, The Observatories, Madingley Road, Cambridge CB3 OHA, United Kingdom}

\newcommand{\nue}{$\nu_e$}
\newcommand{\nua}{${\bar{\nu}}_e$}
\newcommand{\nux}{$\nu_x$}
\newcommand{\msun}{$M_\odot$}
\newcommand{\model}[1]{\texttt{#1}}

\begin{document}

\preprint{APS/123-QED}

\title{Multipolar neutrino radiation in binary neutron star mergers: \\
Angular structure, rotational variability, and \\
implications for electron eraction\\
}

\author{Kutay Arınç Çokluk}
\email{kutay.arinc.cokluk@ege.edu.tr}
\affiliation{\ege}
\affiliation{\egeinst}
\affiliation{\egravity}

\author{Albino Perego}
\affiliation{\unitn}
\affiliation{\infn}

\author{Federico Maria Guercilena}
\affiliation{\unitn}
\affiliation{\infn}

\author{Kadri Yakut}%
\affiliation{\ege}
\affiliation{\egravity}
\affiliation{\ioa}

\date{Received 19 August 2026; accepted 20 October 2026; published 16 November 2026}

\begin{abstract}
The angular structure and temporal variability of neutrino emission from binary neutron star mergers are characterized using fully general-relativistic simulations with energy-integrated M1 neutrino transport across a representative set of equations of state, total masses and mass ratios. The main analysis is restricted to merger remnants that do not collapse to a black hole within the simulated time intervals. The angle-dependent neutrino energy flux is extracted on a spherical surface outside the remnant and decomposed into spherical harmonics to quantify its multipolar content and evolution. Following the initial post-merger transient, the neutrino radiation flux approaches an axisymmetric configuration dominated by a strong quadrupolar component, producing persistent polar flux enhancement and equatorial suppression due to torus shadowing. The dipolar contribution remains subdominant, indicating the absence of sustained one-sided emission. The degree of anisotropy increases with mass asymmetry and for softer equations of state, reflecting the compactness and morphology of the remnant--disk system. Superimposed on this time-averaged geometry, coherent azimuthal modulations associated with the $m=1$ mode are identified. The associated half frequency, $f_{\mathrm{dom}}/2\simeq0.6$--$0.7\,\mathrm{kHz}$, is compared with local rotation rates in the simulated remnant--disk systems, suggesting a possible dynamical association between the remnant structure and neutrino variability. Finally, we quantify how the same quadrupole-dominated radiation geometry induces a latitude-dependent equilibrium electron fraction. The polar material is driven close to the neutrino-equilibrium target, whereas equatorial material remains systematically more neutron rich and below equilibrium.
\end{abstract}


\maketitle


\section{\label{sec:introduction}Introduction}

Binary neutron star (BNS) mergers constitute one of the most extreme environments in relativistic astrophysics, coupling strong-field gravity, supranuclear-density matter, neutrino radiation, and multimessenger observables within a single dynamical event \cite{Shibata2019,Radice2020,Baiotti2017}. The detection of GW170817 and its electromagnetic counterparts, GRB170817A and AT2017gfo, established these mergers as progenitors of short gamma-ray bursts and primary sites of rapid neutron-capture ($r$-process) nucleosynthesis \cite{Abbott2017c, Abbott2017d, Arcavi2017,Chornock2017,Cowperthwaite2017, Coulter2017,Drout2017,Evans2017,Goldstein2017,Hallinan2017,
Kasliwal2017,MurguiaBerthier2017,Nicholl2017,Rosswog2018,
Smartt2017,SoaresSantos2017,Savchenko2017,Tanvir2017,
Tanaka2017,Troja2017,Villar2017,Waxman2018,Kasliwal2022,
Waxman2019}. While gravitational waves constrain bulk binary properties and the nuclear equation of state (EoS) \cite{Abbott2017, Abbott2019a}, the kilonova emission probes the thermodynamic and compositional structure of the ejecta (see, e.g., \citep{Cowan2021,Perego2021} and references therein). Neutrino transport provides the essential physical link between remnant dynamics and these electromagnetic observables \cite{Radice2018, Foucart_2023}.

Following merger, the system either forms a long-lived massive neutron star (MNS) or collapses to a black hole, possibly surrounded by a hot accretion torus \cite{Shibata2011, Bernuzzi2016}. If a MNS forms, its strongly nonaxisymmetric early post-merger dynamics produces intense kilohertz gravitational-wave emission during the first $\sim10$--$20$~ms after merger. This emission carries away energy and angular momentum and contributes to the damping of non-axisymmetric remnant structures. Dense nuclear matter, heated up to tens of MeV scale temperatures, generates intense neutrino emission through charged-current reactions and thermal pair processes \cite{Rosswog2003, Sekiguchi2011, Cusinato_2022}. These neutrinos regulate the relative abundance of neutrons and protons, quantified by the electron fraction, $Y_e$, of the surrounding matter and thereby determine whether lanthanide-rich or lanthanide-poor ejecta dominate the kilonova emission, see e.g. \cite{Metzger2010b, Kasen2017, Perego2017b, Wanajo2014,Metzger2014,Martin2015,Foucart2016LowMass,
Sekiguchi2016,Fujibayashi2017,Radice2018,Fujibayashi2020,
Just2021,Just2022,Radice_2023}. Driven by these multimessenger implications, a vast body of recent general relativistic BNS mergers simulations has extensively explored the binary parameter space, characterizing how different mass ratios, EoSs, and remnant fates dictate the macroscopic ejecta dynamics \cite{Cokluk_2024, Cokluk2023, Camilletti_2022, Combi_2023, Nedora_2021, Just_2022, Fernandez_2022, Zappa_2023, Radice_2023, Padillagay_2024, Radice_2024, Richers_2024, Pajkos_2025, Bernuzzi_2025, Foucart_2024}. To capture the crucial microphysics linking these dynamics to actual observables, modern general-relativistic calculations increasingly rely on two-moment transport schemes, including energy-integrated M1 implementations developed or applied by several independent groups, e.g., \cite{Foucart2016,Shibata2011,Sekiguchi2015,Radice_2022,
Musolino2024,Schianchi2023,Wen2026,Espino2024,Curtis2024,Musolino2025,Neuweiler2026, Foucart2016LowMass,Foucart2015,Cheong2023}. Complementary Monte Carlo transport calculations provide higher-fidelity information about the neutrino energy and angular distributions, e.g., \cite{Foucart2018,Foucart_2020,Richers2015, Miller2019a,Miller2019b,Foucart2021,Foucart2023MC}. Beyond the transport algorithm itself, the resulting neutrino signal and outflow properties can also depend on the adopted interaction rates and reaction set, e.g., \cite{Foucart_2024,Chiesa2025,Rath2026,FoucartRath2026}.

Global, i.e. angle-integrated, neutrino luminosities and mean energies have been investigated since early leakage and flux-limited-diffusion calculations \cite{Rosswog2003, Sekiguchi2011, Dessart2009}. More recent general-relativistic studies have explored the dependence of the global neutrino signal on the equation of state, binary parameters, remnant lifetime, and neutrino-transport treatment e.g., \cite{Sekiguchi2015,Sekiguchi2016,Foucart2016LowMass,
Foucart2016,Vincent2020,Cusinato_2022,Radice_2022,Zappa_2023}. 
A robust outcome of these analyses is that the early postmerger remnant emits neutrinos at total luminosities of several $10^{53}\,\mathrm{erg\,s^{-1}}$, with $\bar{\nu}_e$ typically the most luminous species. 
Hotter and more compact remnants generally produce stronger emission, while the dependence on mass ratio is less universal. Systematic surveys find reduced luminosities for strongly tidally disrupted asymmetric systems, whereas other transport calculations find only weak mass ratio dependence for the electron flavors \cite{Vincent2020, Cusinato_2022}.
For long-lived remnants, neutrino cooling becomes the dominant energy-loss channel after the initial gravitational-wave-dominated phase, and the early $\bar{\nu}_e$ excess decreases as the remnant evolves \cite{Radice_2023}.
The mean energies show a stable flavor ordering, $\langle E_{\nu_x}\rangle > \langle E_{\bar{\nu}_e}\rangle > \langle E_{\nu_e}\rangle$, and change less significantly with binary parameters than the luminosities. Their quantitative values nevertheless retain some dependence on the transport prescription and adopted interaction rates, especially for heavy-lepton neutrinos, e.g., \cite{Foucart_2020,Zappa_2023,Foucart_2024,Cheong2024}.

The angular dependence of the radiation flux at large distance from the remnant is less completely quantified, even though its qualitative geometry is well established.
The emerging radiation originates from strongly inhomogeneous, species- and energy-dependent decoupling regions extending from the dense remnant to the optically thinner disk, e.g., \cite{Endrizzi_2020}. Since neutrino absorption strongly depends on the local neutrino fluxes, the angular distribution of neutrino irradiation directly influences the spatial distribution of $Y_e$ and therefore the viewing-angle dependence of kilonova light curves, including the blue and red components observed in GW170817-like events \cite{Finstad2018, Cokluk2025}. A geometry-resolved characterization of neutrino emission is therefore essential for establishing a quantitative link between remnant dynamics, ejecta composition, and viewing-angle dependent electromagnetic observables.
Pioneering studies realized early that the neutrino emission is preferentially directed along the rotation axis, so that a pole-on observer would infer a larger luminosity than an equatorial observer \cite{Rosswog2003}.
Axisymmetric radiation-hydrodynamics calculations subsequently compared polar and equatorial fluxes explicitly, finding enhanced high-latitude fluxes and energy deposition, while later neutrino-transport simulations showed that the polar regions provide the main escape route for neutrinos trapped in the dense remnant and disk \cite{Dessart2009, Foucart2016LowMass, Vincent2020}. Energy-integrated M1 simulations likewise found that the radiation field is stronger along the rotation axis and that the disk shields lower latitudes from direct irradiation by the massive neutron star \cite{Radice_2022}. Monte Carlo benchmarks further demonstrated that approximate moment closures can be least accurate precisely in the polar regions, motivating a quantitative characterization of the full angular radiation field \cite{Foucart2018}. Related studies of neutrino-driven winds further demonstrated that stronger polar irradiation produces systematically higher electron fractions at high latitudes than in lower-latitude outflows, e.g.,
\cite{Perego2014,Martin2015,Fujibayashi2017}.

Beyond axisymmetric anisotropy, merger remnants frequently develop nonaxisymmetric hydrodynamic activity, including long-lived $m=1$ spiral modes \cite{Bernuzzi2020, Koeppel2019}. These one-armed spiral structures modify density and temperature distributions within the inner remnant and disk and may imprint coherent variability on the neutrino emission, enabling a direct comparison between rotational activity and radiation variability.
Furthermore, the angular structure of neutrino emission has gained renewed importance in the context of fast flavor instabilities. Recent studies demonstrate that electron lepton number crossings and rapid flavor conversion are highly sensitive to the relative angular distributions of neutrinos and antineutrinos, e.g., \cite{Nagakura2023,Froustey2024,Grohs2024,Qiu_2025a}. Because flavor evolution depends explicitly on angular moments of the radiation field, a detailed multipolar mapping of the radiation flux  constitutes a necessary first step toward identifying where such instabilities may arise in merger environments. Flavor conversion may in turn modify neutrino absorption, ejecta composition, and nucleosynthesis, e.g., \cite{Wu2017,George2020,Qiu_2025b}.

In this work, we go beyond previous polar--equatorial comparisons by resolving the neutrino energy flux over the full extraction sphere and by quantifying its time-dependent angular structure with a spherical-harmonic decomposition. This approach separates the direction-averaged component from the dominant axisymmetric quadrupole and from nonaxisymmetric modes, enabling a quantitative characterization of both the persistent polar--equatorial contrast and its azimuthal variability. To quantify this angular structure, we decompose the logarithm of the radial neutrino energy flux into spherical harmonics. In this representation, the monopole ($\ell=0$) describes the angular mean of the logarithmic flux, while the dipole ($\ell=1$) and quadrupole ($\ell=2$) characterize its large-scale angular contrasts. Such a decomposition provides a compact and quantitative description of deviations from isotropy, isolates the leading low-order moments governing the radiation geometry, and enables a direct assessment of their temporal evolution.
Additionally, to relate more explicitly the neutrino fluxes to the ejecta properties, we compare the simulated angular distribution of $Y_e$ with a neutrino-equilibrium target, $Y_{e,\mathrm{eq}}$, thereby separating the composition toward which neutrino irradiation drives the material from the degree to which the ejecta actually reaches that target before weak freeze-out.

This analysis is applied to the neutrino emission emerging from fully general relativistic BNS merger simulations performed using the Einstein Toolkit with energy-integrated M1 transport. 
A suite of binaries spanning different total masses, mass ratios, and equations of state (DD2 and SFHo) is analyzed. The neutrino radiation field is extracted on a spherical surface outside the remnant and decomposed into spherical harmonics, enabling a quantitative determination of its multipolar content and temporal evolution. By resolving the angular and dynamical structure of neutrino emission beyond direction-averaged quantities, this work establishes a quantitatively controlled multipolar framework that connects remnant hydrodynamics, neutrino transport, and multimessenger observables within a single self-consistent description.

Three principal results are established. First, a few tens of milliseconds after merger the neutrino energy flux robustly approaches an axisymmetric configuration dominated by a strong quadrupolar component, producing persistent polar enhancement and equatorial suppression whose amplitude increases with mass asymmetry and EoS softness. Second, coherent $m=1$ azimuthal variability is identified and compared with rotation profiles calculated directly from the simulated fluid. The comparison is intended to investigate a possible connection between neutrino variability and local rotational dynamics. Third, the resulting angle-dependent neutrino irradiation provides physically grounded inputs for viewing-angle dependent kilonova modeling and for future investigations of flavor instabilities in merger remnants. We further confirm that the flux quadrupole translates into a robust polar--equatorial composition imprint: the equilibrium electron fraction is systematically larger toward the poles, while the simulated equatorial electron fraction remains below its neutrino-equilibrium value at late times.
Although our analysis provides a solid and quantitative explanation of the electron fraction distribution inside the ejecta, the detailed nucleosynthesis outcome of the BNS merger simulations used in this work is analyzed in a separate, dedicated paper \citep{Loffredo2026}.

The paper is organized as follows. Section~\ref{sec:methods} describes the numerical setup, including the general-relativistic hydrodynamics framework, energy-integrated M1 neutrino transport, and the binary sample. Section~\ref{sec:neutrino_luminosity_time_evolution} establishes the time evolution of the angle-integrated luminosities and mean energies. Section~\ref{sec:angular_distribution} then examines the angular distribution of the energy flux and its polar--equatorial contrast. Section~\ref{sec:spherical_harm_decomp} presents the multipolar decomposition, the low-order hierarchy, and the nonaxisymmetric $m=1$ modulation together with its connection to remnant rotation. Section~\ref{sec:influence_electron_fraction} discusses the latitude dependence of $Y_{e,\mathrm{eq}}$ and compares the simulated $Y_e$ with its neutrino-equilibrium target. Section~\ref{sec:conclusions} summarizes the conclusions.

\begin{table*}[htbp]
  \centering
  \caption{\label{tab:sim_summary}
  Summary of the BNS models analyzed in this work. For each configuration we list the model name, chirp mass ($\mathcal{M}_c$), equation of state (EoS), mass ratio ($q=M_2/M_1$), total mass ($M_{\rm tot}$), initial ADM mass, merger time ($t_{\rm merger}$), and the final simulation time ($t_{\rm end}$), which is given relative to the merger time. The disk mass and amount of the ejected matter are given in the last two columns.
  }
  
  \setlength{\tabcolsep}{3.5pt} 
  \renewcommand{\arraystretch}{1.2} 
  \small 

  \begin{tabular}{@{}lccccccccc@{}}
    \toprule
    \textbf{Simulation name}
      & \textbf{$M_\mathrm{chirp}$}
      & \textbf{$EoS$}
      & \textbf{$q$}
      & \textbf{$M$}
      & \textbf{$M_\mathrm{ADM}$}
      & \textbf{$t_\mathrm{mer}$}
      & \textbf{$t_\mathrm{end}$} 
      & \textbf{$M_\mathrm{disk}^\mathrm{end}$} 
      & \textbf{$M_\mathrm{ejec}^\mathrm{end}$} \\
      & \textbf{($M_\odot$)}
      & 
      & 
      & \textbf{($M_\odot$)}
      & \textbf{($M_\odot$)}
      & \textbf{($\rm{ms}$)} 
      & \textbf{($\rm{ms}$)} 
      & \textbf{($M_\odot$)} 
      & \textbf{($10^{-3} M_\odot$)} \\ 
    \midrule
    DD2\_M25960\_q100      & 1.13 & DD2  & 1.00 & 2.5960 & 2.5720 & 10.179 & 62.96 & 0.223 & 4.740 \\
    \midrule
    DD2\_M26048\_q085      & 1.13 & DD2  & 0.85 & 2.6048 & 2.5810 & 10.084 & 55.25 & 0.273 & 4.389 \\
    \midrule
    DD2\_M26469\_q070      & 1.13 & DD2  & 0.70 & 2.6469 & 2.6230 & 9.174  & 61.82 & 0.295 & 6.683 \\
    \midrule
    DD2\_M28700\_q100      & 1.25 & DD2  & 1.00 & 2.8700 & 2.8410 & 7.878  & 81.02 & 0.173 & 4.341 \\
    \midrule
    DD2\_M28860\_q085      & 1.25 & DD2  & 0.85 & 2.8860 & 2.8570 & 7.832  & 66.82 & 0.249 & 4.752 \\
    \midrule
    DD2\_M29325\_q070      & 1.25 & DD2  & 0.70 & 2.9325 & 2.9030 & 7.419  & 66.41 & 0.271 & 5.690 \\
    \midrule
    SFHo\_M25960\_q100      & 1.13 & SFHo & 1.00 & 2.5960 & 2.5720 & 11.588 & 74.66 & 0.188 & 3.691 \\
    SFHo\_M25960\_q100\_HR  & 1.13 & SFHo & 1.00 & 2.5960 & 2.5720 & 12.377 & 51.17 & 0.178 & 4.321 \\
    \midrule
    SFHo\_M26048\_q085      & 1.13 & SFHo & 0.85 & 2.6048 & 2.5810 & 11.294 & 50.75 & 0.185 & 5.500 \\
    SFHo\_M26048\_q085\_HR  & 1.13 & SFHo & 0.85 & 2.6048 & 2.5810 & 12.337 & 22.13 & 0.190 & 3.820 \\
    \bottomrule
  \end{tabular}
\end{table*}

\section{Methods}
\label{sec:methods}

\subsection{Simulation sample and numerical methods}
\label{sec:sample_numerical_methods}
A set of ten BNS system models corresponding to eight distinct physical configurations has been simulated, with total initial masses in the range $M \in [2.60, 2.93]$\msun and mass ratios spanning $q \in [0.70, 1.00]$, covering both equal-mass and unequal-mass configurations. The sample is organized around two target chirp masses, $\mathcal{M}_c \simeq 1.13$\msun and $\mathcal{M}_c \simeq 1.25$\msun, which bracket the chirp mass of GW170817 ($\mathcal{M}_c \approx 1.19$\msun). This choice, complementary to the well-studied GW170817 case, is intended to probe how the chirp mass impacts the postmerger dynamics, in particular the remnant lifetime and associated neutrino emission characteristics.
The main sample analysed below consists of configurations in which no black hole forms during the intervals studied. The conclusions concerning sustained neutrino emission and its late-time angular structure therefore refer to surviving neutron-star remnants. An illustrative comparison with a rapidly collapsing configuration is presented separately in Appendix~\ref{app:sfho_remnant_outcome}.

Initial conditions  were constructed from irrotational, quasi-circular BNS configurations computed using the \textsc{LORENE} spectral solver \citep{Gourgoulhon_2001, LORENEcode}. The neutron stars were initially positioned at a coordinate separation of 45 km, corresponding to approximately three–four orbits prior to merger. The initial parameters of the simulated BNS systems, along with selected postmerger quantities, are summarized in Table \ref{tab:sim_summary}. 

We perform fully general relativistic simulations of BNS mergers using the open-source \textsc{Einstein Toolkit} infrastructure\citep{ETKcode, Loffler_2012}. Spacetime evolution is solved by the \textsc{CTGamma} thorn \citep{Pollney_2011,Reisswig_2013}, which employs the Z4c constraint damping scheme \citep{Bernuzzi_2010}, the $1+\log$ slicing condition, and the integrated Gamma-driver shift condition. The metric fields are evolved using fourth-order finite-difference operators, with Kreiss–Oliger dissipation applied to ensure the nonlinear stability of the evolution. The evolution of matter is handled by the general relativistic hydrodynamics module \textsc{WhiskyTHC} \citep{Radice_2012,Radice_2013,Radice_2014}, which implements flux-conservative, high-resolution shock-capturing schemes. A fifth-order monotonicity-preserving reconstruction method \citep{Suresh_1997} is used in combination with an HLLE  approximate Riemann solver \citep{Harten_1983,Einfeldt_1988}. The method of lines is employed to couple the hydrodynamic evolution to the spacetime dynamics, enabling consistent time integration across all evolved variables. Time integration is carried out using a third-order Runge–Kutta (RK3) scheme\citep{Runge_1895, Kutta_1901}, with a fixed Courant–Friedrichs–Lewy factor of 0.075. 
A low-density artificial atmosphere is imposed throughout the computational domain, with a floor density set to $6 \times 10^{3}\, \mathrm{g\, cm^{-3}}$ to ensure numerical stability.

We employ the \textsc{Carpet} adaptive mesh refinement driver \citep{Carpetcode}, utilizing seven levels of nested Cartesian grids centered on the neutron stars during inspiral and on the remnant following merger. For all our models, the finest refinement level achieves a spatial resolution of $\Delta x \approx 246{\rm m} $, which is then considered as our baseline resolution. All the corresponding simulations are evolved for at least 50ms post merger. In a few selected cases, we also evolved the binary with $\Delta x \approx 185{\rm m} $. These latter, more refined, models are labeled as high resolution, \model{HR}. The computational domain extends to $\pm 1512 {\rm km}$ in each spatial direction. To reduce computational cost, reflection symmetry is imposed across the equatorial plane.

\subsection{Equation of state}
\label{sec:eos}
Thermal and compositional evolution of the fluid is modeled using finite-temperature, composition-dependent microphysical EoSs. We adopt the HS(DD2) \citep{Hempel_2012}, in the following simply indicated as DD2, and SFHo\citep{Steiner_2013} nuclear EoSs to describe the behavior of matter inside the merging neutron stars and in the merger remnant. SFHo is often considered as a ``soft" nuclear EoS, predicting a radius of approximately $11.9\,{\rm km}$ and a dimensionless tidal deformability of $\Lambda_{1.4} \approx 334$ for a $1.4$ \msun neutron star. These macroscopic properties are in agreement with the constraints derived from the GW170817 event, which estimated a radius of $11.9\pm1.4\, {\rm km}$ and placed an upper limit on the tidal deformability at $\Lambda_{1.4}<800$ \citep{Abbott2017}. Conversely, the DD2 EoS exhibits a ``stiffer" behavior at supranuclear densities, supporting a maximum mass of $2.42$\msun for a cold, beta-equilibrated and nonrotating configuration.
For a $1.4$ \msun neutron star, DD2 yields a larger radius of $13.2\, {\rm km}$ and a tidal deformability of $\Lambda_{1.4} \approx 702$. While DD2 approaches the upper bound of the GW170817 $90\%$ confidence interval, it remains a viable and widely used baseline model for exploring the phenomenology of stiff nuclear matter.

\subsection{Neutrino transport and microphysics}
\label{sec:neutrino_transport_microphysics}
Neutrino transport is treated using the moment-based M1 scheme via the \textsc{THC\_M1} \citep{Radice_2022} module implemented in the \textsc{WhiskyTHC} code. This formalism approximates the full Boltzmann transport equation by evolving the lowest two energy-integrated moments of the neutrino distribution function: the energy density ($E$) and the momentum density flux \textbf{($F^i$)} 
. The transport equations are cast into a 3+1 conservative form and are fully coupled to the hydrodynamics and spacetime evolution through the divergence of the stress-energy tensor.

To close the system of moment equations, an analytic closure relation is required to express the pressure tensor in terms of $E$ and \textbf{$F^i$}
. We adopt the Minerbo closure, which assumes a maximum entropy distribution for radiation. This closure provides an interpolation between the optically thick regime (diffusion limit, where the Eddington factor is $1/3$) and the optically thin regime (free-streaming limit, where the Eddington factor approaches $1$). The transition between these regimes is governed by the evolved flux factor, ensuring causality and stability in both opaque and transparent regions.

Regarding microphysics, we include three neutrino flavors, modeled as three independent neutrino species: $\nu_e$, $\bar{\nu}_e$ and $\nu_x$, where the latter is a collective species for muonic and tauonic (anti)neutrinos. The interaction rates used to compute the neutrino source terms and opacities comprise the following processes:
\begin{enumerate}
    \item \textbf{Beta (charged-current) processes:} Electron capture ($e^- + p \rightarrow n + \nu_e$) \citep{Bruenn_1985} and positron capture ($e^+ + n \rightarrow p + \bar{\nu}_e$) \citep{Bruenn_1985}, along with their inverse absorption reactions. These are the dominant drivers of composition ($Y_e$) evolution.
    \item \textbf{Neutrino-pair (thermal) processes:} Electron-positron pair annihilation ($e^- + e^+ \rightarrow \nu + \bar{\nu}$) \citep{Ruffert_1997} and nucleon-nucleon bremsstrahlung ($N + N \rightarrow N + N + \nu + \bar{\nu}$) \citep{Burrows_2006}, which dominate the production of heavy-lepton neutrinos ($\nu_x$).
    \item \textbf{Scattering:} Elastic scattering of neutrinos on nucleons ($\nu + N \rightarrow \nu + N$) \citep{Ruffert_1997} and coherent scattering on heavy nuclei ($\nu + A \rightarrow \nu + A$) \citep{Shapiro_1983}, which serve as the primary opacity sources determining the neutrino diffusion time.
\end{enumerate}
For more details about the rate implementation, see \citep{Radice_2022} and references therein.

\subsection{Postanalysis methods}
\label{sec:pa_methods}

\subsubsection{Gravitational waves}
\label{sec:pa_methods_gw}

To analyze the simulations and facilitate comparisons across models, the time of merger is determined for each run. To this end, the gravitational wave signal is extracted at a coordinate radius of approximately $295\,\mathrm{km}$ from the center of mass of the system. The Newman–Penrose scalar $\Psi_4$ is computed using the \textsc{WeylScal4} module of the \textsc{Einstein Toolkit}, and decomposed into spin-weighted spherical harmonics ( $_{-2}Y_{lm}\left(\theta,\phi\right)$) using the \textsc{Multipole} module. The scalar $\Psi_4$ is related to the second time derivatives of the gravitational wave polarizations via the expansion:
\begin{equation}
    \label{eq:newman-penrose_scalar}
    \Psi_4= \ddot{h}_+ -i\ddot{h}_\times=\sum_{l=2}^{\infty}\sum_{m=-l}^{l}{\psi_4^{lm}\left(t,r\right)}\ _{-2}Y_{lm}\left(\theta,\phi\right).
\end{equation}
The \textsc{Python} library \textsc{Kuibit}\citep{kuibit_2021} is used to compute the gravitational wave polarization modes $h_+$ and $h_\times$ from the Newman–Penrose scalar $\Psi_4$, according to the relation:

\begin{equation}
    \label{eq:gw_polarization}
    h_+^{\ell m}(r,t) - i h_\times^{\ell m}(r,t) = \int_{-\infty}^{t} du \int_{-\infty}^{u} dv\, \psi_4^{\ell m}(r,v).
\end{equation}
Since the gravitational wave signals are extracted at a finite coordinate distance from the system’s center of mass, propagation delays must be accounted for to determine the physical merger time. To this end, the simulation time is corrected by computing the retarded time. 
\begin{equation}
    \label{eq:retarded_time}
    t_{\mathrm{ret}} = t - R - 2M_0 \ln{\left( \frac{R}{2M_0} - 1 \right)},
\end{equation}
where $R$ is the extraction radius and $M_0$ is the initial ADM mass of the system.

\subsubsection{Disk and ejecta mass}
\label{sec:pa_methods_disk_ejecta_mass}

The disk mass is defined as the bound material with rest-mass density below $10^{13}\,\mathrm{g\,cm^{-3}}$, and is computed via a general relativistic volume integral of the conserved baryon mass density over the computational domain:
\begin{equation}
    \label{eq:disk_mass}
    M_{\mathrm{disk}} = \int_{\Omega} \sqrt{\gamma} \, \rho W \, d\Omega. 
\end{equation}
In this expression, $\rho$ denotes the baryon rest-mass density, $W$ is the fluid Lorentz factor, and $\sqrt{\gamma}$ is the determinant of the spatial three-metric.

The total ejected mass is evaluated as the cumulative mass flux of unbound material through a spherical surface located at approximately \(295\,\mathrm{km}\), using the Bernoulli criterion, \(-h u_t > 1\). The values measured at the end of the simulations are reported in Table~\ref{tab:sim_summary}.

\subsubsection{Neutrino quantities}
\label{sec:pa_methods_neutrino}

We start by characterizing the neutrino emission through angle integrated quantities.
In particular, we consider the time evolution of the total energy luminosity, number luminosity, and mean energy, extracted for each neutrino species at the edge of our computational domain.
The total luminosity is computed according to
\begin{equation}
    \label{eq:nu_luminostiy}
    L = \int_{\Omega} F_{\nu,r} \, dA = \int_{\Omega} F_{\nu,r} \, r^2 \, d\Omega
\end{equation}
where $F_{\nu,r}$ represents the radial energy flux through a spherical surface located at approximately $295\,\mathrm{km}$. 
The neutrino number luminosity is computed using an expression analogous to Eq.~(\ref{eq:nu_luminostiy}), but using the number flux.
The mean energy for each species is then obtained as the ratio of its radial energy luminosity to its radial number luminosity.

To analyze the angular dependence of the neutrino emission, we consider extraction surfaces and spherical angular coordinates ($\theta$,$\phi$) on them. More specifically, we discretize the azimuthal angle in $N_{\phi} = 93$, equally spaced bins. For the polar angle $\theta$, we consider $N_{\theta}= 51 $ equally spaced bins. Such a grid enables sufficient angular resolution to capture anisotropies in the neutrino emission.
To directly compare fluxes from the polar to equatorial regions with the total luminosity, we introduce polar-dependent, azimuthally averaged, isotropized luminosities:
\begin{equation}
    \label{eq:iso_luminosity}
    L_{\mathrm{iso}}(\theta_i) = 2 \Delta\phi \, r^2 \sum_{j} F_{\nu,r}(\theta_i, \phi_j).
\end{equation}
In this expression, $F_{\nu,r}(\theta_i, \phi_j)$ denotes the neutrino flux at the angular grid coordinates $(\theta_i, \phi_j)$, $\Delta\phi$ is the azimuthal angular resolution, and $r$ is the extraction radius. Physically, this quantity represents the equivalent isotropic luminosity that a distant observer located at viewing angle $\theta$ would infer, assuming the entire emission surface radiates at the flux density measured at that specific latitude. This angular characterization of neutrino luminosity is crucial for understanding the latitudinal variations in neutrino irradiation and, consequently, for interpreting the observed anisotropy in kilonova emission.

The electron fraction of matter is regulated by weak interactions inside the simulations. In optically thin conditions, if neutrino fluxes are sufficiently intense or the exposure time is sufficiently long, it evolves toward values close to an equilibrium value, which is approximated by 
\citep{Qian_1996}.
\begin{equation}
    \label{eq:ye_asym}
    Y_{e,{\rm eq}}
    \approx \left(1+{L_{\bar\nu_e}\over L_{\nu_e}}{\epsilon_{\bar\nu_e}-2\Delta
    +1.2\Delta^2/\epsilon_{\bar\nu_e}\over 
    \epsilon_{\nu_e}+2\Delta+1.2\Delta^2/
    \epsilon_{\nu_e}}\right)^{-1}
\end{equation}
where $L_{\nu_e}$ and $L_{\bar{\nu}e}$ are the energy luminosities, $\epsilon_{\nu_e}$ and $\epsilon_{\bar{\nu}_e}$ are parameters proportional to the mean energies of electron neutrinos and antineutrinos, respectively, and $\Delta = 1.293\,\mathrm{MeV}$ is the neutron–proton mass difference. Following \citep{Qian_1996}, $\epsilon_\nu$ denotes the spectral moment $\langle E_\nu^2\rangle/\langle E_\nu\rangle$, not the ordinary mean energy. Because the present extraction provides the ratio of energy to number luminosity rather than the second energy moment, we adopt the approximate spectral-shape factor $\epsilon_\nu/\langle E_\nu\rangle\simeq1.2$ proposed by \citet{Qian_1996} in their Sec.~5.1, following their Eqs.~(65a)--(65b). We apply this prescription consistently to all models, times, and angular directions. The factor is an assumed spectral approximation, not a directly measured spectral moment of the present simulations. In particular, it was derived under the assumption of steady-state beta equilibrium in a neutrino capture-dominated regime, where electron and positron captures are subdominant. It assumes that the rates of $\nu_e + n \rightarrow p + e^-$ and $\bar{\nu}_e + p \rightarrow n + e^+$ interactions dominate the evolution of $Y_e$, and that neutrino and antineutrino absorption are the primary processes setting the composition of the ejecta.

For the analysis in Sec.~\ref{sec:influence_electron_fraction}, $Y_{e,\mathrm{eq}}$ is evaluated over
5-ms bins from the binned neutrino fields. Regional means and contrasts are weighted by the detector surface element; unless otherwise stated, polar regions satisfy $|\mathrm{\theta}|\geq60^\circ$ and equatorial regions satisfy $|\mathrm{\theta}|\leq15^\circ$.

\subsubsection{Spherical harmonic analysis and rotational profile}
\label{sec:pa_methods_rot_profile}

The angular characterization of the neutrino emission is achieved
by expanding the logarithm of the radial neutrino energy flux,
$f_{\nu,r}(\theta,\phi,t) \equiv
\log_{10}{\left(F_{\nu,r}(\theta,\phi,t)\right)}$,
according to:
\begin{equation}
    \label{eq:shd_expansion}
    f_{\nu,r}(\theta,\phi,t)
    = \sum_{\ell,m} a_{\ell m}(t)\,Y_{\ell m}(\theta,\phi),
\end{equation}
where $Y_{\ell m}(\theta,\phi)$ are the real, $4\pi$-normalized spherical-harmonic basis functions, $a_{\ell m}$ are the corresponding expansion coefficients, $\theta$ and $\phi$ denote the polar and azimuthal angles, 
and $\ell$ and $m$ are the degree and order of the harmonic, respectively. 
We use the logarithm of the flux to characterize relative angular contrasts over a broad dynamic range. The logarithmic transformation compresses the contrast between bright and faint directions and describes multiplicative variations as additive angular structure. Thus, the coefficients $a_{\ell m}$ quantify the contribution of each spherical harmonic mode to the angular structure of the logarithm of the flux and are computed via:
\begin{equation}
    \label{eq:spd_modes}
    a_{\ell m}(t) = \frac{1}{4\pi} \int_{\Omega} f_{\nu,r}(\theta,\phi,t) \, Y_{\ell m}(\theta,\phi) \, d\Omega \, .
\end{equation}
These coefficients and the associated angular power spectrum, 
\(P_\ell \equiv \sum_m |a_{\ell m}|^2\), are extracted from the simulation output using the \textsc{SHTOOLS} library, which implements numerical routines for spherical harmonic transforms on equiangular grids. The detailed formulation can be found in
\citep{Wieczorek_2018}.
For consistency with the notation used below, we denote individual expansion coefficients by $C_{\ell m}$ in the text and figures, with $C_{\ell m}\equiv a_{\ell m}$. In particular, $C_{\ell0}$ denotes the axisymmetric component, which depends only on the polar angle, whereas coefficients with $m\neq0$, such as $C_{11}$, describe azimuthal variations.

Geometrically, each coefficient $a_{\ell m}$ represents the projection of the angular distribution of the logarithm of the flux onto the corresponding spherical harmonic mode $(\ell, m)$. Lower-degree modes, such as $\ell = 0$ and $\ell = 1$, capture large-scale angular features. The $\ell = 0$ mode represents the isotropic (monopole) component, while the $\ell = 1$ modes characterize dipolar asymmetries. Specifically, the $(\ell, m) = (1, 1)$ mode corresponds to a dipole aligned with the $x$ axis, $(1, 0)$ corresponds to a dipole along the $z$ axis (the polar direction), and $(1, -1)$ corresponds to a dipole along the $y$ axis.

The angular velocity profile of the massive neutron star in the equatorial plane during the postmerger phase is computed according to

\begin{equation}
    \label{eq:omega}
    \Omega = \alpha v^\phi - \beta^\phi
\end{equation}
where $\alpha$ is the lapse function, $v^\phi$ is the azimuthal component of the fluid three velocity, and $\beta^\phi$ is the corresponding component of the shift vector \citep{Hanauske_2017}. These are defined in terms of Cartesian coordinates as:
\begin{equation}
    \label{eq:azimuthal_3vel}
    v^\phi = \frac{x v^y - y v^x}{x^2 + y^2 + z^2},
\end{equation}
 \begin{equation}
    \label{eq:azimuthal_shiftvector}
     \beta^\phi = \frac{x \beta^y - y \beta^x}{x^2 + y^2 + z^2}.
 \end{equation}

To characterize the differential rotation of the postmerger remnant, we construct radial rotation profiles on the equatorial plane. At each postmerger time $t_i$, the $z=0$ Cartesian-grid data are remapped onto a uniform polar grid $(r,\phi)$. 
The polar grid covers the full azimuthal range, $\phi\in[-\pi,\pi]$, with 120 radial and 360 azimuthal grid points.
We first compute the azimuthally averaged angular velocity,
\begin{equation}
\label{eq:azimuthally_averaged_omega}
    \left\langle \Omega \right\rangle_{\phi}(r,t_i)
    =
    \frac{1}{2\pi}
    \int_{-\pi}^{\pi}
    \Omega(r,\phi,t_i)\,d\phi \, ,
\end{equation}
which in the following will often be presented in terms of rotation frequency, $\left\langle \Omega \right\rangle_{\phi}/2 \pi$, and expressed in kHz.

For each simulation time, $\left\langle \Omega \right\rangle_{\phi}(r,t_i)$ is paired with the corresponding azimuthally averaged rest-mass-density profile,
\begin{equation}
    \label{eq:azimuthally_averaged_density}
    \left\langle \rho \right\rangle_{\phi}(r,t_i)
    =
    \frac{1}{2\pi}
    \int_{-\pi}^{\pi}
    \rho(r,\phi,t_i)\,d\phi \, .
\end{equation}
This provides a radial, parametric relation between $\left\langle \rho \right\rangle_{\phi}$ and $\left\langle \Omega \right\rangle_{\phi}$, which is used to identify the rotational structure of the remnant, including differentially rotating regions and shear layers.

To suppress short-time-scale variations in the matter rotation profile, we additionally compute a time- and azimuthally averaged rotational frequency following the procedure of \citet{Hanauske_2017}. For a time window of width $\Delta t$, we define
\begin{equation}
    \label{eq:time-azimuthally-averaged-omega}
    \frac{\overline{\Omega}(r,t)}{2\pi}
    =
    \frac{1}{\Delta t\,(2\pi)^2}
    \int_{t-\Delta t/2}^{t+\Delta t/2}
    \int_{-\pi}^{\pi}
    \Omega(r,\phi,t')\,d\phi\,dt' \, .
\end{equation}
Here, $\overline{\Omega}(r,t)/(2\pi)$ denotes the rotation frequency obtained by averaging $\Omega$ over the full azimuthal range and over the temporal interval $\Delta t$. We adopt $\Delta t=1\,\mathrm{ms}$ for the time-dependent matter--neutrino comparisons.

\section{Temporal evolution of the angle-integrated luminosities and mean energies}
\label{sec:neutrino_luminosity_time_evolution}

\begin{figure*}
\includegraphics[width=1\linewidth]{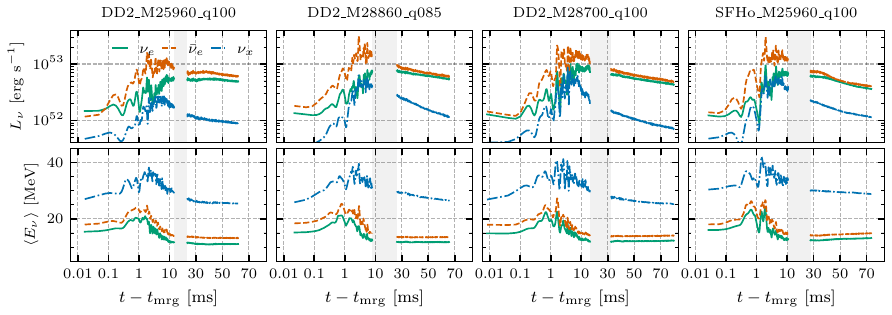}
  \caption{  \label{fig:2fig_lum_mean_afo_time}Time evolution of the neutrino emission for four representative BNS merger models. Each column shows one simulation. \emph{Top row:} total energy luminosities;  \emph{bottom row:} mean neutrino energies. Colors denote different flavors such as green, orange and blue correspond to \nue, \nua and \nux, respectively. Shaded areas correspond to time windows in which simulation data were missing. The time axis is initially ($0.01-10~\rm{ms}$) in logaritmic scale and later in linear scale.
  } 
\end{figure*}

The temporal evolution of neutrino luminosities and mean energies in BNS
mergers reflects the coupled effects of merger dynamics, general relativity, and neutrino--matter microphysics \cite{Sekiguchi2011, Foucart2016LowMass, Fujibayashi2017, Radice_2023}. Thus a nontrivial influence of the system's mass ratio, total initial mass, and nuclear EoS is expected \cite{Cusinato_2022, Endrizzi_2020}.

During the cold inspiral, neutrino luminosities are negligible  ($L_\nu \lesssim 10^{51}\,\rm erg\,\rm s^{-1}$) as tidal heating remains insufficient to trigger significant emission. Upon merger, the violent collision and core fusion convert bulk kinetic energy into thermal energy, causing luminosities to surge to peak values of a few $10^{53}\, \rm erg\,s^{-1}$ within the first 5 ms postmerger. 
For models that form a long-lived massive neutron star, as the ones presented in Fig.~\ref{fig:2fig_lum_mean_afo_time}, the merger peak is followed by an early postmerger oscillatory phase with several well-defined luminosity maxima (typically 3--4) over $5$--$20$~ms, associated with quasiradial bounces and oscillations of the remnant core. At $t>20$~ms, the luminosities enter a secular decline.
During the simulation, a robust flavor hierarchy emerges in the luminosity curves: $L_{\bar{\nu}_e} >  L_{\nu_e} >  L_{\nu_x}$.  The dominance of electron antineutrinos, which is particularly evident in the early postmerger phase,  
is driven by the fact that the merger remnant is initially extremely neutron-rich and shock-heated, favoring positron captures on free neutrons over electron captures on protons.

The neutrino mean energies exhibit a different hierarchy: $\langle E_{\nu_x}\rangle > \langle E_{\bar{\nu}_e}\rangle > \langle E_{\nu_e}\rangle$. This ordering is dictated by the depth of the respective neutrinospheres; heavy-lepton neutrinos decouple in deeper, hotter layers due to their lower opacities, while electron neutrinos have the highest absorption opacities and decouple in cooler, more distant regions. Postmerger, these mean energies remain remarkably stationary because the thermodynamic conditions at the decoupling surfaces evolve slowly during the accretion disk's lifetime.

The influence of the initial binary parameters on the emission characteristic is clearly distinguishable. Comparing the DD2 equal-mass models (first and third columns), an increase in the total binary mass from $M=2.60~$\msun to $M=2.87~$\msun leads to a notable enhancement in the peak luminosities. This is attributed to the higher compactness and larger temperatures achieved in the more massive neutron star remnant. Furthermore, the impact of the nuclear EoS is evident when contrasting the DD2 and SFHo models for fixed mass parameters ($M=2.60$ \msun, $q=1.0$). The SFHo model (rightmost column), which employs a softer EoS, yields a slightly harder neutrino spectrum, particularly for the \nux~species. This spectral hardening suggests that the softer EoS leads to a more compact remnant with higher central temperatures, thereby shifting the neutrino 
distribution to higher energies relative to the stiffer DD2 model. Despite these differences in the spectral hardness, the overall luminosity evolution remains qualitatively robust across the explored EoS variations, dominated primarily by the cooling timescale of the remnant.

\section{Angular distribution}
\label{sec:angular_distribution}

\subsection{Distribution over the detector sphere}
\label{sec:detector_sphere_distribution}

\begin{figure*}
  \centering
  \includegraphics[width=0.9\linewidth]{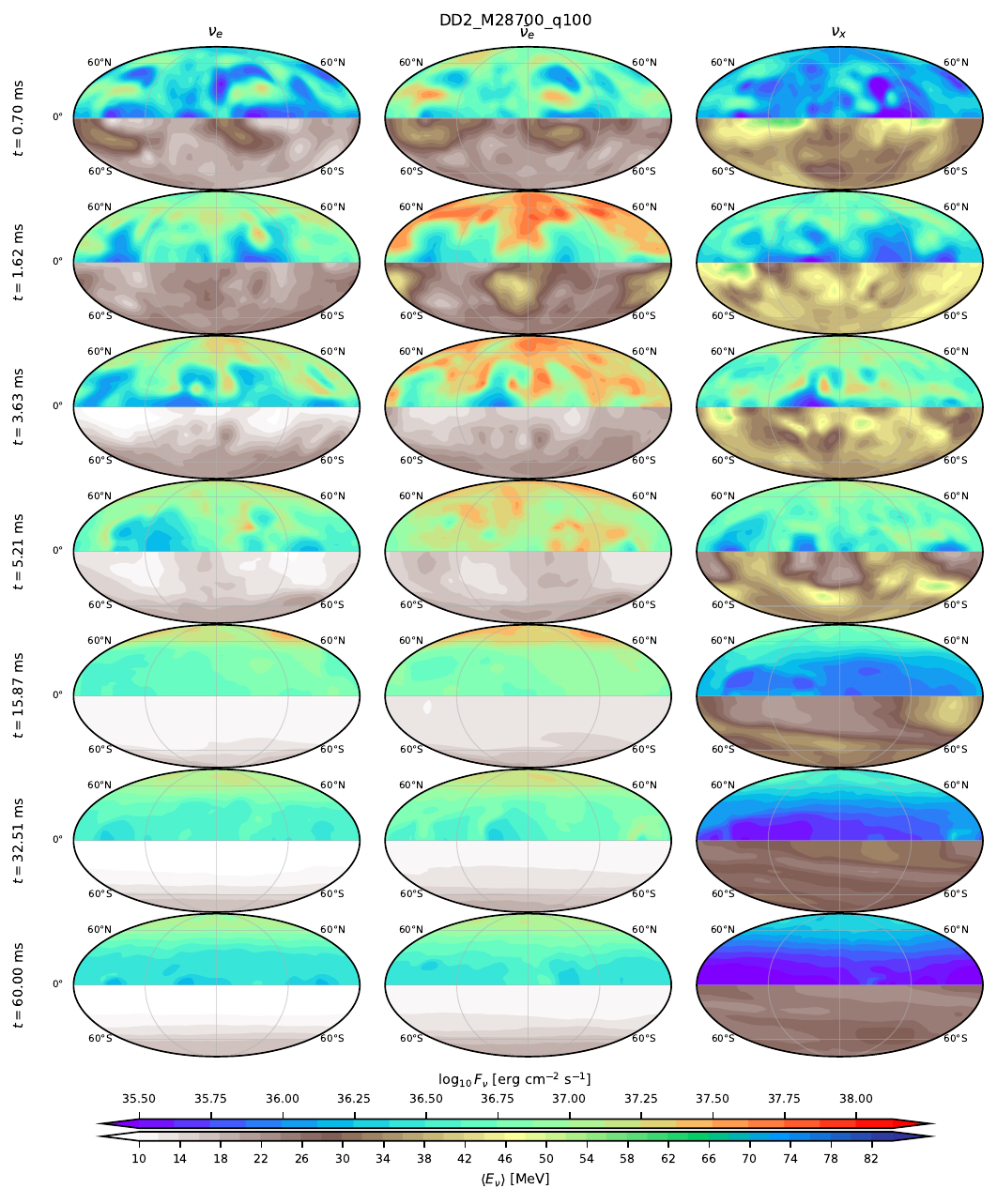}
  \caption{
  \label{fig:angular_dist_nu_DD2_M2870_q100}
Angular distributions of neutrino energy flux and mean energy for the \texttt{DD2\_M28700\_q100} model, extracted on a sphere at $R\simeq295$~km. Rows show the postmerger times $t=0.70$, $1.62$, $3.63$, $5.21$, $15.87$, $32.51$, and $60.00$~ms from top to bottom; columns show $\nu_e$, $\bar{\nu}_e$, and $\nu_x$ from left to right. Each panel is a Mollweide projection. The northern displayed hemisphere shows $\log_{10}F_r$ in $\mathrm{erg\,cm^{-2}\,s^{-1}}$, while the southern displayed hemisphere shows the mean energy $\langle E_\nu\rangle$ in MeV. The split-hemisphere layout is solely a compact visualization of two quantities and does not represent a physical north--south comparison. The upper and lower shared colorbars give the energy-flux and mean-energy scales, respectively, and are common to all times and species. The mean-energy maps use colour levels separated by $2\,\mathrm{MeV}$, with a common energy scale across the panels to resolve the late-time angular variations while preserving direct comparisons between species and epochs.
}
\end{figure*}

Figure~\ref{fig:angular_dist_nu_DD2_M2870_q100} shows Mollweide maps of the neutrino emission extracted on the sphere at $R\simeq295$~km for the \texttt{DD2\_M28700\_q100} model. Rows correspond to selected postmerger times, $t=0.70$, $1.62$, $3.63$, $5.21$, $15.87$, $32.51$, and $60.00$~ms, while columns show $\nu_e$, $\bar{\nu}_e$, and $\nu_x$, respectively. Longitude spans $-180^\circ$ to $+180^\circ$, centered at $0^\circ$, and latitude spans $-90^\circ$ to $+90^\circ$. For compactness, the northern hemisphere contains $\log_{10}F_r$ in $\mathrm{erg\,cm^{-2}\,s^{-1}}$, whereas the southern one contains the mean energy $\langle E_\nu\rangle$ in MeV. This split-hemisphere representation is a visualization convention and does not indicate a physical north--south asymmetry.

The energy-flux maps reveal a fast-growing and later persistent large-scale polar--equatorial morphology. At all times, the electron-flavor fluxes are enhanced at high latitudes and suppressed around the equator. 

At early times, $t\lesssim5$~ms, both the flux and mean-energy maps exhibit substantial azimuthal structure. Localized bright and dim regions are especially apparent in the electron-flavor fluxes, although the development of a reduced equatorial flux is already visible in all flavors, while the main differences between $\bar{\nu}_e$ and the other species is the boosted emission at high latitudes. The mean-energy distributions are likewise nonaxisymmetric and their latitudinal structure is initially flavor dependent. The electron-flavor mean energies progressively develop lower values near the equator and higher close to the poles.
At the same time, the $\nu_x$ mean energy retains a rather irregular structure, with high values over a broad range of latitudes and does not show an equally uniform polar enhancement at all early snapshots.

By $t\simeq15$--$60$~ms, the angular distributions become smoother and increasingly dominated by their latitude dependence. The electron-flavor flux maps approach a nearly axisymmetric configuration with bright polar regions and a continuous dim equatorial belt. Over the same interval, $\nu_e$ and $\bar{\nu}_e$ display a clear mean-energy increase away from the equatorial region. The $\nu_x$ mean energy remains substantially larger than those of the electron flavors, while its angular contrast is comparatively weaker and evolves more gradually.

Overall, the emission evolves from an azimuthally structured early postmerger state toward a smoother, pole-focused radiation field. The robust features are the persistent polar--equatorial anisotropy of the energy flux, the dominance of $\bar{\nu}_e$ in the total energy flux, and the stable spectral hierarchy with $\nu_x$ carrying the largest mean energies. These angular and flavor-dependent properties provide the neutrino-radiation context for the latitude-dependent weak-interaction processing and equilibrium electron-fraction trends discussed in Sec.~\ref{sec:influence_electron_fraction}.

\begin{figure*}
  \centering
  \includegraphics[width=1\linewidth]{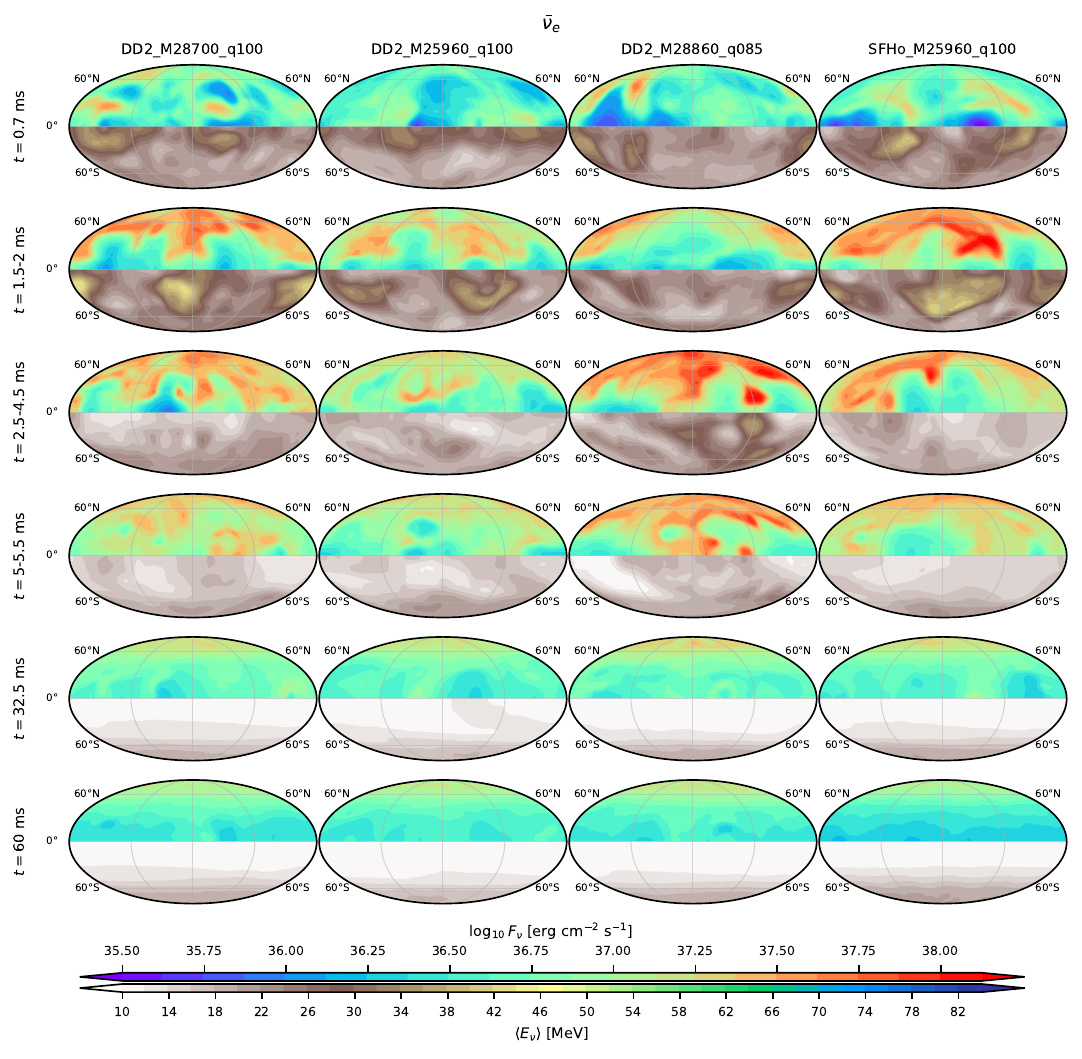}
  \caption{
    \label{fig:angular_dist_nue_comparison}
    Same as Fig.~\ref{fig:angular_dist_nu_DD2_M2870_q100}, but for electron-antineutrino emission, $\bar{\nu}_e$, and for four representative BNS merger models. Columns show, from left to right, \texttt{DD2\_M28700\_q100}, \texttt{DD2\_M25960\_q100}, \texttt{DD2\_M28860\_q085}, and \texttt{SFHo\_M25960\_q100}. Rows show snapshots at $t\simeq0.7$~ms, $1.5$--$2$~ms, $2.5$--$4.5$~ms, $5$--$5.5$~ms, $32.5$~ms, and $60$~ms. The exact early-time snapshot varies slightly between models within the indicated intervals.
    }
\end{figure*}

\begin{figure*}
  \centering
  \includegraphics[width=1\linewidth]{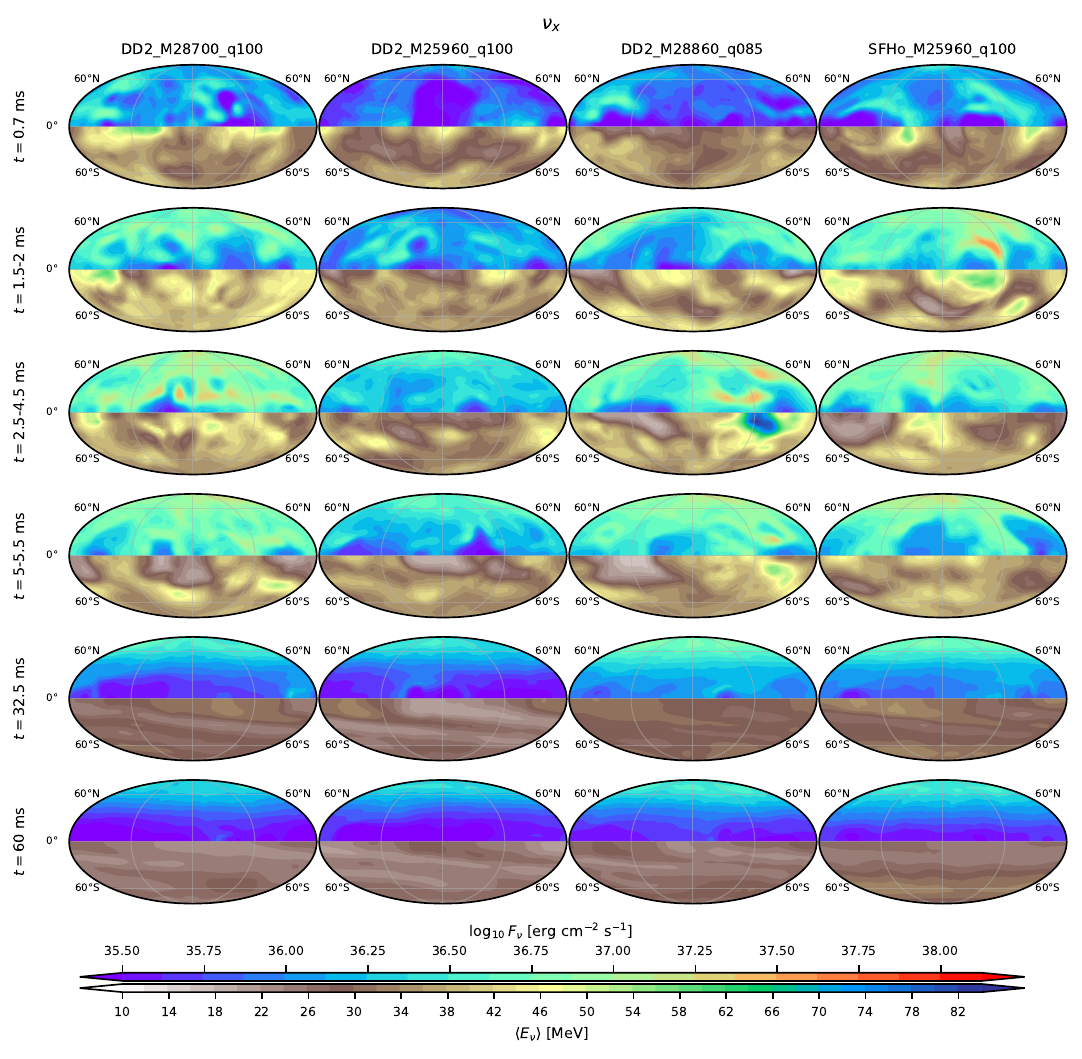}
  \caption{
    \label{fig:angular_dist_nux_comparison}
    Same as Fig.~\ref{fig:angular_dist_nue_comparison}, but for heavy-lepton
    neutrinos, $\nu_x$. The northern displayed hemisphere shows
    $\log_{10}F_r$ in $\mathrm{erg\,cm^{-2}\,s^{-1}}$, and the southern displayed
    hemisphere shows $\langle E_{\nu_x}\rangle$ in MeV.
    }
\end{figure*}

Figures~\ref{fig:angular_dist_nue_comparison} and \ref{fig:angular_dist_nux_comparison} extend the single-model comparison of Fig.~\ref{fig:angular_dist_nu_DD2_M2870_q100} to all four BNS merger models presented in Fig.\ref{fig:2fig_lum_mean_afo_time}, but focusing on the $\bar{\nu}_e$ and $\nu_x$ channels, respectively. The first two columns compare equal-mass DD2 binaries with $M_{\rm tot}\simeq2.87\,M_\odot$ and $2.60\,M_\odot$, respectively. The comparison of the first and third columns probes the effect of a reduced mass ratio at nearly fixed total mass, whereas the second and fourth columns isolate the EsS dependence for equal-mass binaries. The common color scales and common angular representation permit a direct comparison of both the absolute emission level and its angular structure.

The early postmerger maps retain a pronounced azimuthal structure in both flavors and for all models. By $t\simeq2.5$--$5.5$~ms, the energy flux has developed a clearer high-latitude enhancement and equatorial suppression in most models, and this morphology remains the dominant large-scale feature at $t\simeq32.5$ and $60$~ms. The mean-energy evolution is more nuanced. For $\bar{\nu}_e$, a low-energy equatorial region and hotter high-latitude emission become increasingly clear after the first few milliseconds. The $\nu_x$ radiation remains substantially harder than the electron-flavor
radiation, but its mean-energy anisotropy is weaker and is not uniformly monotonic at the earliest times. Thus, the robust common feature of these maps is the polar--equatorial flux anisotropy, while the detailed mean-energy morphology retains a stronger flavor and time dependence.

At fixed mass ratio, the more massive equal-mass DD2 model is brighter in both $\bar{\nu}_e$ and $\nu_x$ during the first several milliseconds and generally exhibits higher mean energies. The difference in absolute $\bar{\nu}_e$ flux decreases substantially by $t\simeq32.5$--$60$~ms, showing that total mass primarily controls the early emission level rather than enforcing a uniform late-time flux offset. The comparison between \texttt{DD2\_M28700\_q100} and \texttt{DD2\_M28860\_q085} instead shows that the mass ratio mainly changes the time-dependent azimuthal morphology
and the detailed polar--equatorial contrast. In particular, the unequal-mass configuration displays more localized early-time structure, while its late-time contrast is comparable to, and in several electron-antineutrino snapshots stronger than, that of the equal-mass model. This behavior is consistent with a radiation geometry modified by tidal asymmetry and nonaxisymmetric accretion, although the maps alone do not identify a unique density or opacity origin.

The EoS comparison is especially clear in the $\nu_x$ channel. Relative to \texttt{DD2\_M25960\_q100}, the \texttt{SFHo\_M25960\_q100} model generally shows a brighter and harder $\nu_x$ radiation and retains larger
high-latitude mean energies at late times. For $\bar{\nu}_e$, the SFHo model is also brighter during the early postmerger evolution, while at late times its most robust signature is a stronger angular contrast and higher polar mean energies rather than a uniformly larger flux at every latitude. These trends are consistent with the more compact remnant expected for the softer EoS, which modifies both the neutrino-decoupling geometry and the relative importance of polar escape and equatorial attenuation. The angular maps therefore motivate the quantitative polar-angle and multipole analyses presented below.

\begin{figure*}
  \centering
  \includegraphics[width=1\linewidth]{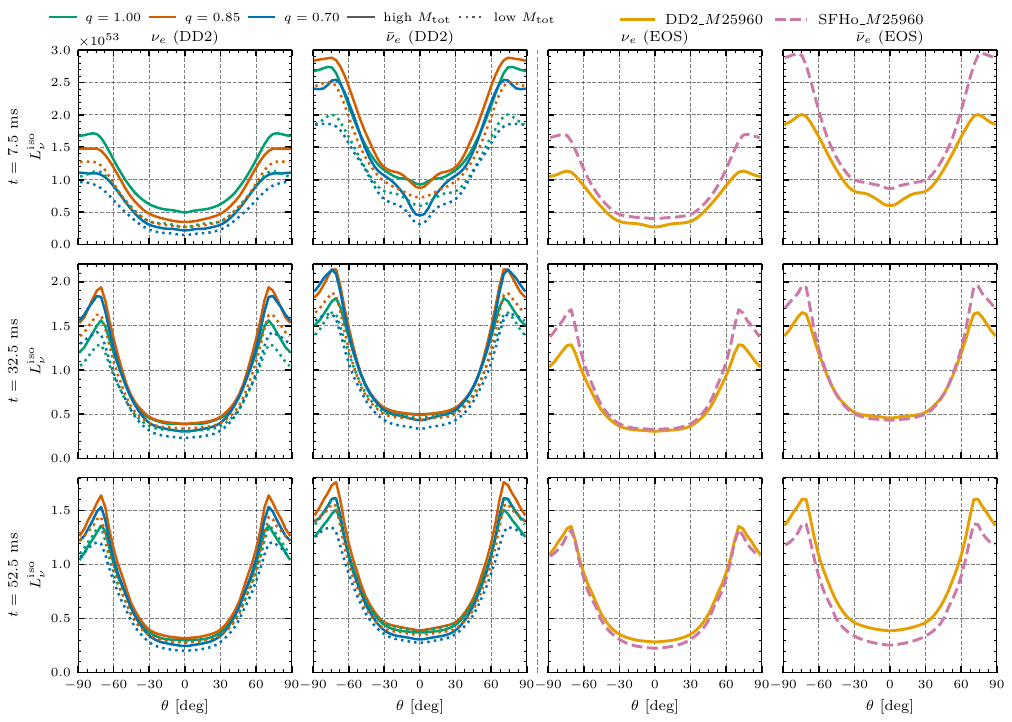}
  \caption{
  \label{fig:isolum_afo_theta_3times_M_q_eos}
  Angular dependence of the isotropic neutrino luminosities for six DD2 binary neutron star merger models with high and low total masses and mass ratios $q=0.70-1.00$ (\textit{first and second columns}) and for two BNS merger models \model{DD2\_M25960\_q100} and \model{SFHo\_M25960\_q100} 
  (\textit{third and fourth columns}) to reveal the impact of the EsS. Each column shows the luminosity as a function of the polar angle $\theta$ at three representative post-merger times: $t = 7.5\,\mathrm{ms}$ (top row), $t = 32.5\,\mathrm{ms}$ (middle row), and $t = 52.5\,\mathrm{ms}$ (bottom row). Within each pair, the left column reports electron neutrinos (\nue) and the right column electron antineutrinos (\nua).
  }
\end{figure*}

\begin{figure*}
  \centering  \includegraphics[width=1\linewidth]{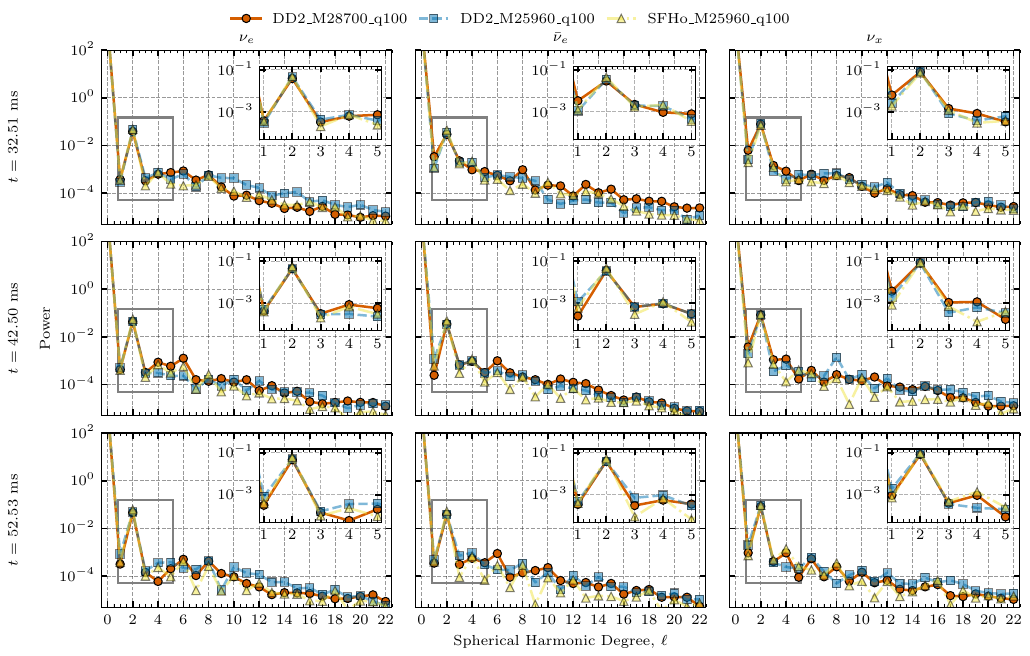}
  \caption{
  \label{fig:EF_power_afo_l}
  Spherical harmonic power spectrum of the \nue{}, \nua{} and \nux{} logarithm of radial energy flux on the extraction sphere as a function of degree $l$, shown at $t=32.5$, $42.5$, and $52.5$ ms. Three models are plotted: \model{DD2\_M28700\_q100} (orange circles), \model{DD2\_M25960\_q100} (blue squares), and \model{SFHo\_M25960\_q100} (yellow triangles). The zoomed inset box highlights low degree values.
  }
 
\end{figure*}

\subsection{Luminosity distribution as a function of polar angle}
\label{sec:polar_angle_distribution}

To better quantify the dominant anisotropy degree along the polar direction in the neutrino emission, in Fig.~\ref{fig:isolum_afo_theta_3times_M_q_eos} (left two panels) we present the isotropic luminosities of electron neutrinos and antineutrinos as a function of the polar angle $\theta$ for six DD2 binaries with high and low total masses and mass ratios $q=0.70{-}1.00$, evaluated at $t=7.5$, $32.5$, and $52.5\mathrm{ms}$. In all cases, the angular distributions display a characteristic U-shape: luminosities are suppressed around the equator ($\theta\simeq0^\circ$) and enhanced toward the poles ($\theta\simeq\pm90^\circ$). This pattern is shared by both \nue and \nua flavors, consistent with earlier multidimensional studies of neutrino emission from binary neutron
star merger remnants \cite{Rosswog2003,Dessart2009,Foucart2016LowMass,Vincent2020}. 
The absolute scale is dominated by \nua's, as expected from the luminosity hierarchy discussed in Sec.~\ref{sec:neutrino_luminosity_time_evolution}, while \nue's exhibit systematically larger polar--equator (P/E) ratios, as their decoupling surfaces are located further out in regions where equatorial opacity remains higher.

For a fixed mass ratio, the overall luminosity level scales with the total binary mass. High-mass systems achieve significantly larger isotropic luminosities at early times ($t=7.5\,\mathrm{ms}$), typically exceeding their low-mass counterparts by $\sim30\%$. However, the anisotropy quantified by P/E ratios is only weakly affected. The luminosity excess observed in the high-mass models decrease at later stages ($t=32.5{-}52.5\,\mathrm{ms}$), as their remnants cool and contract more rapidly, leading to steeper declines in emission.

The degree of angular anisotropy is primarily governed by the mass ratio, while it also grows with post-merger time in equal-mass and moderately asymmetric models.
Equal-mass binaries ($q=1.0$) show moderate contrasts, with $\mathrm{P/E}\approx 2.5{-}2.8$ (for \nua) and $\approx 2.9{-}3.5$ (for \nue) at $t=7.5\,\mathrm{ms}$, rising to $\approx 3.5$ and $\approx 4.1$ by $t=52.5\,\mathrm{ms}$. Asymmetric systems ($q=0.70$) are markedly more anisotropic: already at $t=7.5\,\mathrm{ms}$ they reach $\mathrm{P/E}\approx 4.0$ (for \nua) and $\approx 4.7$ (for \nue), with values persisting around $4.2$ and $5.2$ at later times. Intermediate ratios ($q=0.85$) occupy an intermediate regime ($\mathrm{P/E}\approx3.7{-}4.4$). These trends reflect the remnant morphology: tidal disruption in asymmetric mergers produces thicker accretion disks that enhance equatorial baryon loading, thereby suppressing equatorial emission and sharpening the polar funnels.

The comparison, visible in the right two panels of Fig.~\ref{fig:isolum_afo_theta_3times_M_q_eos}, between models employing the DD2 and SFHo EoSs for BNS characterized by $M=2.60\,M_\odot$ and $q=1.0$, further illustrates how the EoS also controls anisotropy. At $t=7.5\,\mathrm{ms}$, the P/E ratios are essentially identical ($2.8$ for \nua, $3.5$ for \nue), although SFHo yields higher absolute luminosities. However, differences grow with time. At $t=32.5\,\mathrm{ms}$, SFHo reaches $\mathrm{P/E}=3.7$ (\nua) and $4.3$ (\nue), compared to DD2’s $3.1$ and $3.6$. At later times ($t=52.5\,\mathrm{ms}$), the contrast further widens ($4.5$ vs $3.5$ and $4.9$ vs $4.1$, respectively). Softer EoS remnants are more compact and hotter, enhancing charged-current emission and increasing equatorial optical depths, thereby producing stronger polar focusing of the neutrino flux.

The quantitative polar-to-equatorial ratios are subject to systematic uncertainties associated with the M1 closure. Comparisons with Monte Carlo transport have shown that M1 can overestimate polar emission and produce an artificially strong polar enhancement of the neutrino energy flux \citep{Gizzi_2021,Foucart_2024}. The P/E values reported here should therefore be interpreted as predictions within the adopted transport approximation. The magnitude of this bias has not been calibrated for the present models, so these comparisons do not provide a universal correction factor for our results.

\begin{figure*}
  \centering
  \includegraphics[width=1\linewidth]{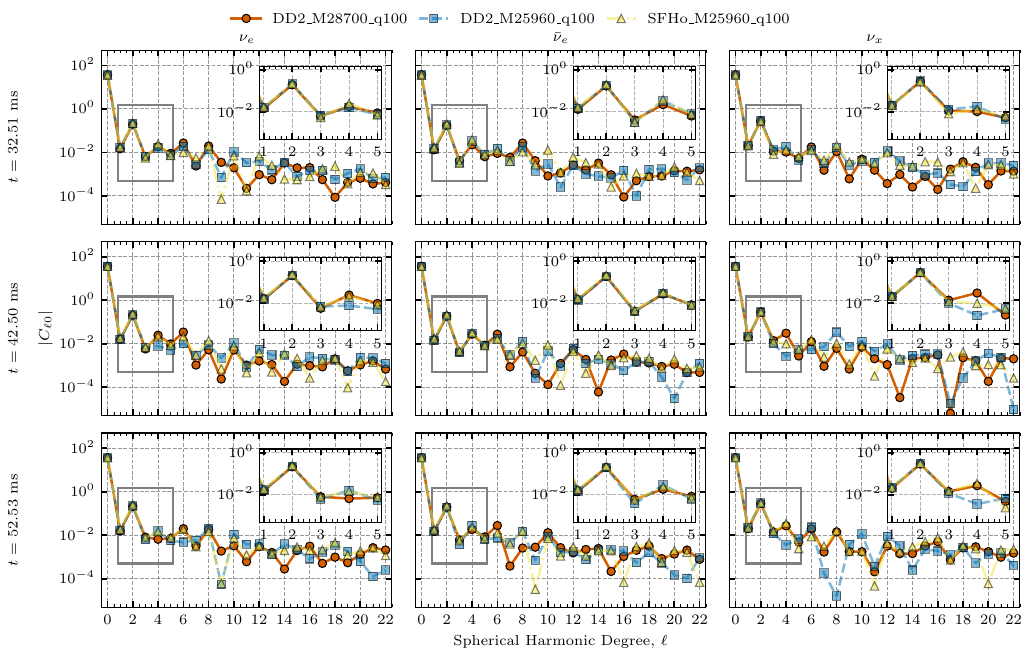}
  \caption{Same as Fig.~\ref{fig:EF_power_afo_l}, but for the amplitude of the axisymmetric ($m=0$) spherical-harmonic coefficients of the \nue{}, \nua{} and \nux{} logarithm of radial energy flux as a function of degree $l$, at $t=32.5$, $42.5$, and $52.5$ ms.}
  \label{fig:EF_C_l0_afo_l}
\end{figure*}


\section{Spherical harmonic decomposition}
\label{sec:spherical_harm_decomp}

To characterize the large-scale angular structure of the neutrino radiation, we have expanded the logarithm of the radial energy flux on the extraction sphere in spherical harmonics and computed the resulting power spectrum 
(see Sec.\ref{sec:pa_methods_rot_profile}). In what follows, we focus on the lowest multipoles \(\ell=0,1,2\), which capture the bulk of the anisotropy at late post-merger times, and we discuss three representative epochs (\(t=32.5,\,42.5,\,52.5\) ms) for the models \model{DD2\_M25960\_q100}, \model{DD2\_M28700\_q100}, and \model{SFHo\_M25960\_q100}. The corresponding coefficients for \nue, \nua, and \nux~are reported in Table~\ref{tab:combined_spectra_panels} (a), and presented in Fig.~\ref{fig:EF_power_afo_l}.

Across all species and models the anisotropic power is dominated by \(\ell=2\), whereas \(\ell=1\) remains at least an order of magnitude smaller. For instance, at \(t=32.5\) ms, \nue~exhibits \(P_2\sim (3.9\text{–}4.8)\times10^{-2}\) while \(P_1\sim (2.9\text{–}3.7)\times10^{-4}\). This hierarchy indicates that the neutrino flux is largely axisymmetric at late times and shaped by the polar–equatorial contrast (consistent with the maps in Sec.~\ref{sec:detector_sphere_distribution} and the angular profiles in Sec.~\ref{sec:polar_angle_distribution}) resulting from the enhanced high latitude emission and the low latitude shadowing by the disk .

At fixed total mass \(M_{\rm tot}=2.60\,M_\odot\), the model characterized by the softer SFHo EoS systematically yields larger \(P_2\) than the DD2 model for all flavors and times. For \nue, \(P_2\) rises from \(4.81\times10^{-2}\) (32.5 ms) to \(6.03\times10^{-2}\) (52.5 ms) in the SFHo case, compared with \(4.71\times10^{-2}\rightarrow4.94\times10^{-2}\) for the DD2 case. Similar trends hold for \nua~and \nux. This behavior is consistent with the more compact and hotter remnant produced in the SFHo model, which increases equatorial opacities and sharpen polar focusing.

Comparing the \model{DD2\_M25960\_q100} and \model{DD2\_M28700\_q100} models allows one to isolate the effects of the total mass. For \nue, \(P_2\) grows modestly in both cases, with a larger relative increase in the high-mass model ($\approx$ 20\% from 32.5 to 52.5 ms) than in the low-mass one ($\approx$ 5\%). For \nua, the high-mass model shows the strongest late-time growth among the DD2 cases. These trends support the picture that more massive remnants remain hotter and more compact for a longer time, sustaining larger polar–equatorial contrasts as the disk cools and thins.

\begin{table*}
\centering
\small
\setlength{\tabcolsep}{1.8pt} 
\renewcommand{\arraystretch}{1.1}
\caption{
\label{tab:combined_spectra_panels}
Temporal evolution of the neutrino radiation properties. The table presents data for $\nu_e$, $\bar{\nu}_e$, and $\nu_x$ at three late postmerger times ($t=32.5, 42.5, 52.5$ ms). (a) Angular power spectra $P_\ell$ obtained from the spherical-harmonic decomposition of the logarithm of the radial neutrino energy flux for representative models during the late postmerger phase, on the extraction sphere at $R\simeq295\,\mathrm{km}$. (b) Axisymmetric coefficients $C_{\ell0}$ of the same logarithmic flux distribution. These coefficients isolate the $m=0$ contributions and therefore quantify the degree of axisymmetry of the radiation field. The strong $\ell=2$ component confirms that the late-time neutrino emission is well described by a monopole plus quadrupole structure. Model abbreviations: DD2-M28 (DD2\_M28700\_q100), DD2-M25 (DD2\_M25960\_q100), SFHo-M25 (SFHo\_M25960\_q100). Values are scaled as indicated in the headers.
}
\begin{tabular}{@{} ll ccc ccc ccc ccc ccc ccc @{}}
\toprule
\multicolumn{20}{c}{\textbf{(a) Power Spectrum Coefficients ($P_\ell$)}} \\
\midrule
 &  & \multicolumn{3}{c}{$P_0 (\times 10^3)$} & \multicolumn{3}{c}{$P_1 (\times 10^{-4})$} & \multicolumn{3}{c}{$P_2 (\times 10^{-2})$} & \multicolumn{3}{c}{$P_3 (\times 10^{-4})$} & \multicolumn{3}{c}{$P_4 (\times 10^{-4})$} & \multicolumn{3}{c}{$P_5 (\times 10^{-4})$} \\
\cmidrule(lr){3-5} \cmidrule(lr){6-8} \cmidrule(lr){9-11} \cmidrule(lr){12-14} \cmidrule(lr){15-17} \cmidrule(lr){18-20}
$t$ (ms) & Model & $\nu_e$ & $\bar{\nu}_e$ & $\nu_x$ & $\nu_e$ & $\bar{\nu}_e$ & $\nu_x$ & $\nu_e$ & $\bar{\nu}_e$ & $\nu_x$ & $\nu_e$ & $\bar{\nu}_e$ & $\nu_x$ & $\nu_e$ & $\bar{\nu}_e$ & $\nu_x$ & $\nu_e$ & $\bar{\nu}_e$ & $\nu_x$ \\
\midrule
32.5 & DD2-M28 & 1.348 & 1.356 & 1.342 & 3.74 & 33.85 & 63.07 & 3.88 & 2.93 & 7.91 & 3.29 & 22.61 & 14.49 & 6.27 & 9.56 & 8.39 & 7.29 & 8.03 & 3.38 \\
     & DD2-M25 & 1.340 & 1.351 & 1.331 & 2.89 & 11.10 & 25.62 & 4.71 & 3.54 & 8.41 & 4.27 & 19.84 & 8.47 & 7.30 & 19.16 & 3.85 & 3.84 & 4.55 & 6.19 \\
     & SFHo-M25 & 1.345 & 1.351 & 1.353 & 3.54 & 12.18 & 19.24 & 4.81 & 4.45 & 7.20 & 2.04 & 17.19 & 12.17 & 7.31 & 21.73 & 3.00 & 2.42 & 3.60 & 3.76 \\
\addlinespace
42.5 & DD2-M28 & 1.345 & 1.351 & 1.335 & 4.28 & 2.48 & 38.08 & 4.42 & 3.59 & 8.39 & 3.22 & 6.68 & 10.98 & 8.66 & 9.52 & 11.71 & 5.86 & 3.17 & 1.71 \\
     & DD2-M25 & 1.342 & 1.351 & 1.328 & 5.07 & 11.97 & 19.64 & 4.40 & 3.41 & 8.18 & 2.99 & 6.73 & 3.59 & 3.11 & 9.93 & 6.37 & 2.34 & 3.10 & 3.49 \\
     & SFHo-M25 & 1.341 & 1.345 & 1.345 & 4.01 & 6.09 & 8.34 & 5.58 & 4.74 & 8.88 & 2.06 & 3.01 & 7.23 & 6.86 & 11.74 & 1.33 & 3.09 & 1.29 & 4.04 \\
\addlinespace
52.5 & DD2-M28 & 1.342 & 1.347 & 1.330 & 3.17 & 3.48 & 9.36 & 4.64 & 3.89 & 8.28 & 1.42 & 3.08 & 4.01 & 0.59 & 5.54 & 9.33 & 2.02 & 3.60 & 0.89 \\
     & DD2-M25 & 1.339 & 1.348 & 1.326 & 8.42 & 4.71 & 19.74 & 4.94 & 3.80 & 9.23 & 1.63 & 7.14 & 3.63 & 3.60 & 9.53 & 2.38 & 3.80 & 2.99 & 2.04 \\
     & SFHo-M25 & 1.335 & 1.338 & 1.339 & 3.84 & 3.95 & 7.19 & 6.03 & 5.21 & 9.36 & 0.99 & 0.90 & 4.78 & 2.30 & 6.71 & 13.78 & 0.93 & 0.68 & 2.52 \\
\midrule[\heavyrulewidth]
\multicolumn{20}{c}{\textbf{(b) Axisymmetric Multipole Coefficients ($C_{\ell 0}$)}} \\
\midrule
 &  & \multicolumn{3}{c}{$C_{00} (\times 10^1)$} & \multicolumn{3}{c}{$C_{10} (\times 10^{-2})$} & \multicolumn{3}{c}{$C_{20} (\times 10^{-1})$} & \multicolumn{3}{c}{$C_{30} (\times 10^{-3})$} & \multicolumn{3}{c}{$C_{40} (\times 10^{-2})$} & \multicolumn{3}{c}{$C_{50} (\times 10^{-3})$} \\
\cmidrule(lr){3-5} \cmidrule(lr){6-8} \cmidrule(lr){9-11} \cmidrule(lr){12-14} \cmidrule(lr){15-17} \cmidrule(lr){18-20}
$t$ (ms) & Model & $\nu_e$ & $\bar{\nu}_e$ & $\nu_x$ & $\nu_e$ & $\bar{\nu}_e$ & $\nu_x$ & $\nu_e$ & $\bar{\nu}_e$ & $\nu_x$ & $\nu_e$ & $\bar{\nu}_e$ & $\nu_x$ & $\nu_e$ & $\bar{\nu}_e$ & $\nu_x$ & $\nu_e$ & $\bar{\nu}_e$ & $\nu_x$ \\
\midrule
32.5 & DD2-M28 & 3.672 & 3.683 & 3.663 & 1.46 & 1.32 & 2.03 & 1.97 & 1.71 & 2.79 & 6.34 & 4.15 & 11.55 & 1.81 & 2.24 & 1.04 & 8.92 & 6.39 & 5.29 \\
     & DD2-M25 & 3.661 & 3.676 & 3.649 & 1.62 & 1.45 & 2.08 & 2.15 & 1.83 & 2.90 & 6.66 & 3.33 & 12.90 & 1.61 & 3.33 & 1.85 & 7.44 & 7.95 & 4.43 \\
     & SFHo-M25 & 3.668 & 3.676 & 3.679 & 1.69 & 1.65 & 2.04 & 2.19 & 2.05 & 2.68 & 5.52 & 3.19 & 8.21 & 2.50 & 3.82 & 1.31 & 7.24 & 7.90 & 6.51 \\
\addlinespace
42.5 & DD2-M28 & 3.668 & 3.676 & 3.655 & 1.59 & 1.48 & 2.05 & 2.09 & 1.88 & 2.89 & 5.82 & 4.18 & 14.02 & 2.47 & 2.77 & 3.08 & 9.88 & 8.46 & 2.76 \\
     & DD2-M25 & 3.664 & 3.676 & 3.645 & 1.57 & 1.43 & 2.13 & 2.10 & 1.84 & 2.85 & 6.98 & 4.05 & 10.01 & 0.78 & 2.93 & 0.27 & 5.24 & 8.11 & 5.33 \\
     & SFHo-M25 & 3.663 & 3.668 & 3.668 & 1.79 & 1.71 & 2.17 & 2.36 & 2.18 & 2.98 & 6.79 & 4.55 & 12.26 & 2.25 & 3.25 & 1.01 & 8.70 & 8.26 & 5.49 \\
\addlinespace
52.5 & DD2-M28 & 3.664 & 3.671 & 3.647 & 1.57 & 1.48 & 2.01 & 2.15 & 1.97 & 2.88 & 8.05 & 5.99 & 14.22 & 0.66 & 1.83 & 2.70 & 7.20 & 8.38 & 4.81 \\
     & DD2-M25 & 3.660 & 3.672 & 3.643 & 1.70 & 1.53 & 2.23 & 2.22 & 1.93 & 3.04 & 6.26 & 3.72 & 11.80 & 1.62 & 2.86 & 0.36 & 5.56 & 6.33 & 6.80 \\
     & SFHo-M25 & 3.655 & 3.659 & 3.660 & 1.85 & 1.77 & 2.15 & 2.46 & 2.28 & 3.06 & 7.76 & 5.65 & 15.13 & 1.47 & 2.41 & 3.52 & 7.60 & 7.11 & 2.44 \\
\bottomrule
\end{tabular}
\end{table*}

Heavy-lepton neutrinos display the largest absolute quadrupole amplitudes at all times (\nux: \(P_2\simeq0.08\text{–}0.09\) between 32.5 and 52.5 ms), with comparatively weak time evolution, consistent with their deeper decoupling and dominant neutral-current interactions. Electron flavors exhibit a stronger temporal increase of \(P_2\), especially for the SFHo model, reflecting their sensitivity to charged-current opacities and the evolving optical-depth structure of the disk–funnel system. 

The increase of $P_2$ with time in the SFHo runs parallels the growth of the polar-to-equatorial luminosity ratio measured from \(L_{\rm iso}(\theta)\) (see Ref.~\ref{sec:polar_angle_distribution}). Likewise, the comparatively steady \(P_2\) for \nux matches its smoother angular maps. The rapid fall-off of power for \(\ell\ge3\) confirms that the emission geometry is controlled by low-order, nearly axisymmetric structure at these late times.

Combining these findings with the results of Secs.~\ref{sec:neutrino_luminosity_time_evolution}--\ref{sec:angular_distribution}, we obtain a coherent scenario: after an early transient, the remnant–disk system settles into a quasi-stationary configuration where polar radiation channels dominate. Softer EoS and larger total mass both enhance compactness and temperature, increasing equatorial opacity; the resulting polar focusing manifests as a strengthened \(\ell=2\) component. The small \(\ell=1\) power and the suppression of higher multipoles indicate mild nonaxisymmetric activity at late times, consistent with the weakening of spiral dynamics and the approach to a steady polar-wind regime.

\begin{figure*}
  \centering
  \includegraphics[width=1\linewidth]{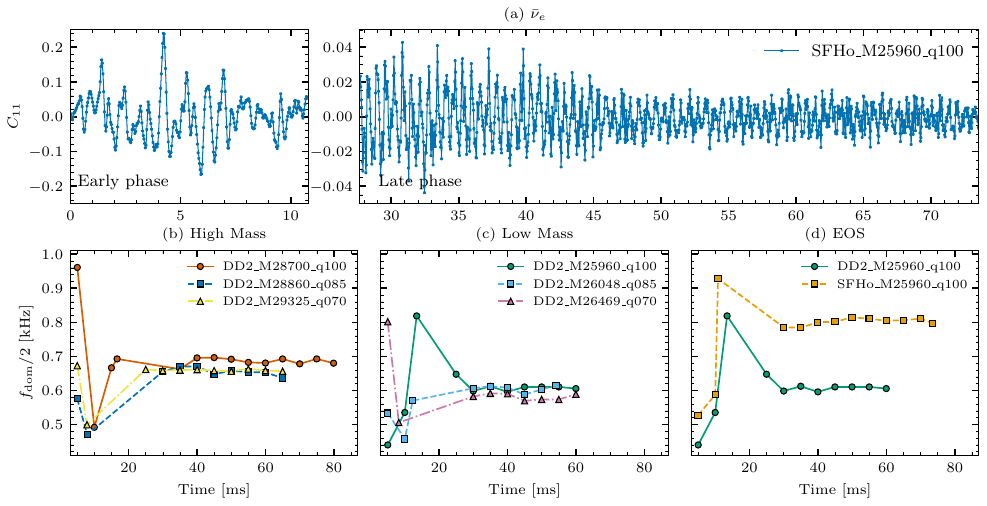}
  \caption{
  \label{fig:EF_c11_domfreq_afo_time}
  (a) Time series of the dipole coefficient $C_{11}$ obtained from the spherical-harmonic decomposition of the logarithm of the radial electron-antineutrino energy flux for the model \model{SFHo\_M25960\_q100}.
  (b) Dominant frequency extracted from 5-ms windowed FFTs of $C_{11}(t)$ for \nua for the high-mass, low-mass (panel c), and EoS comparisons (panel d). We plot half frequency to ease comparison with the rotation rate. After $\sim$30~ms, all models, except SFHo one, converge to a narrow $0.6$--$0.7$~kHz band.
  }
\end{figure*}

Moreover Fig.~\ref{fig:EF_C_l0_afo_l} shows the axisymmetric spherical-harmonic coefficients $C_{\ell0}$ of the logarithm of the radial neutrino energy flux on the extraction sphere as a function of degree $\ell$, for \nue, \nua, and \nux\ at three representative late postmerger times (\(t=32.5,\ 42.5,\ 52.5\) ms) and for the three reference models (DD2\_M28700\_q100, DD2\_M25960\_q100, SFHo\_M25960\_q100). These panels provide a visual overview of how the large scale anisotropy is distributed over the degree \(l\), highlighting the relative importance of the lowest multipoles for each flavor, model, and time.

The numerical values underlying these figures, restricted to the observationally most relevant low orders (\(l\le 5\)), are reported in Table~\ref{tab:combined_spectra_panels} (b) at the same three times (\(t=32.5,\ 42.5,\ 52.5\) ms). In particular, the tables list \(C_{00}\) (monopole), \(C_{10}\) (dipole),\(C_{20}\) (quadrupole), and the next few degrees up to \(C_{50}\), enabling a quantitative reading of the trends
visible in the figures and a direct, model–by–model comparison across EoS and total mass.
The monopole \(C_{00}\), characterizing the angular mean of the logarithmic flux, changes only at the few percent level across models and epochs, so differences in angular structure are not driven by the total flux.
The dominant anisotropic contribution is the quadrupole \(C_{20}\) for all flavors and times. For \nue\ and \nua, the SFHo model systematically yields larger \(C_{20}\) than the two DD2 runs at each epoch (e.g. for \nue: \(2.19,\,2.36,\,2.46\times10^{-1}\) versus \(1.97,\,2.09,\,2.15\times10^{-1}\) for DD2\_M28700\_q100), and \(C_{20}\) increases mildly with time. This trend matches the angular luminosity profiles discussed above: a softer Eos produces a more compact, hotter remnant, enhancing polar emission relative to the equator and thus amplifying the quadrupole.

The dipole \(C_{10}\) remains subdominant for all flavors, with values in the \(\sim10^{-2}\) band and no large temporal excursions, indicating the absence of a strong one–sided brightening at late times. The \(C_{40}\), which is the other subdominant modes, is modest for the electron species, reinforcing a morphology that is well described by a monopole plus a quadrupole, with only gentle higher–order corrections.

Heavy–lepton neutrinos, \nux, share the strong quadrupole but also exhibit a more robust \(l=3\) shoulder. \(C_{30}\) remains at the \(\sim10^{-2}\) level across the inspected epochs (e.g. \(8.21\times10^{-3}\!\to\!1.513\times10^{-2}\) for the SFHo simulation). Although smaller than the quadrupole, this persistent contribution indicates additional latitudinal modulation beyond a simple pole–equator contrast. 

This behavior is consistent with \nux\ decoupling at deeper, hotter layers due to the pair character of the dominant reactions that in our simulations drive their thermal coupling with matter. Combined with quasielastic neutral-current scattering, these reactions preserve a finer angular structure compared to the charged-current-dominated electron flavors.

\begin{figure*}
  \centering
    \includegraphics[width=1\linewidth]{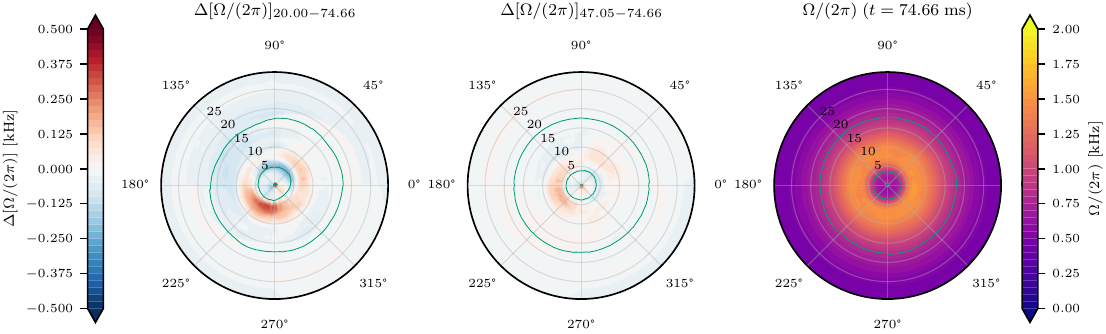}
  \caption{
  \label{fig:omega_in_polar_coord}
  Equatorial angular velocity $\Omega/(2\pi)$ (kHz) in polar coordinates for \texttt{SFHo\_M25960\_q100}. The right panel shows $\Omega/(2\pi)$ at $t-t_{\mathrm{merger}}=74.66$~ms, at the end of the simulation. The middle and left panels display the difference $\Delta\Omega/(2\pi)$ with respect to earlier snapshots at $t-t_{\mathrm{merger}}=47.05$~ms and $t-t_{\mathrm{merger}}=20.00$~ms, respectively, illustrating the secular spin-down of the remnant. The predominantly red shading in the difference maps indicates that the core rotation frequency decreases by $\lesssim 0.2$~kHz over $\sim\!55$~ms, while the outer envelope ($r\gtrsim 20$~km) remains nearly stationary. Green contours mark $\Omega/(2\pi)=0.8$~kHz, cospatial with the inner-disk layers just outside $\rho\simeq10^{13}\;\mathrm{g\,cm^{-3}}$.}
\end{figure*}

Panel (a) of Fig.~\ref{fig:EF_c11_domfreq_afo_time} shows the time series of the dipole coefficient $C_{11}$ obtained from the spherical-harmonic decomposition of the logarithm of the radial electron-antineutrino energy flux for the \texttt{SFHo\_M25960\_q100} model. While the coefficients $C_{\ell0}$ describe the axisymmetric latitude dependence of this flux, $C_{11}$ measures one component of its $m=1$ azimuthal asymmetry.
Immediately after merger, $C_{11}$ exhibits large and irregular oscillations during the dynamical rearrangement of the remnant. At later times, particularly for $t\gtrsim50\,\mathrm{ms}$, the oscillations become more regular. This behavior is consistent with an azimuthal emission pattern rotating at an approximately constant angular speed.

To study this variability, we compute the dominant frequency of the $C_{11}$ coefficient in sliding windows of $\Delta t=5$~ms and follow it in time (Fig.~\ref{fig:EF_c11_domfreq_afo_time}, panels (b)-(d)). The late-time values converge to a narrow band around $0.60$–$0.80$~kHz (we plot half frequency to compare more directly with the remnant rotation rate). Three simple trends appear:
(i) \emph{Total mass:} at fixed mass ratio, DD2 systems characterized by larger masses reach slightly higher plateau frequencies than lower-mass systems, which is consistent with faster spinning, more compact remnants.
(ii) \emph{Mass ratio:} unequal-mass binaries show larger scatter in the first milliseconds, but at late times they approach the same frequency band as the equal-mass cases.
(iii) \emph{EoS:} for equal-mass runs, the SFHo model gives a mildly higher late-time frequency than the DD2 one for the same total mass, in line with a more compact and hotter remnant.

\begin{figure*}
  \centering  \includegraphics[width=1\linewidth]{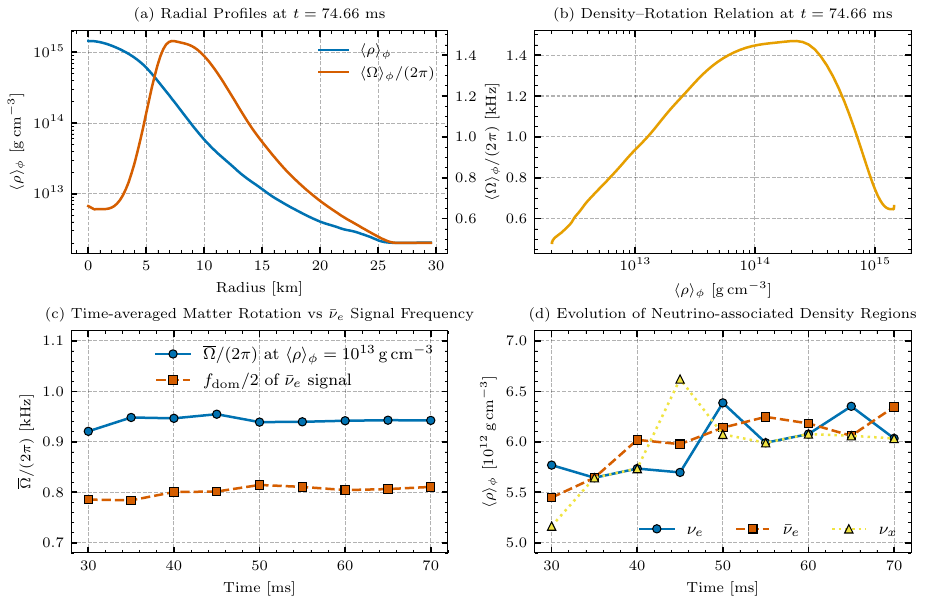}
  \caption{
  \label{fig:EF_density_radial_profile_signal_matter_freq_density_region}
  (a) Radial profiles at a representative late time: the azimuthally averaged rest-mass density $\langle\rho\rangle_{\phi}$ (blue solid line) and the corresponding azimuthally averaged rotational frequency $\langle\Omega\rangle_{\phi}/(2\pi)$ (orange solid line). (b) The associated parametric relation between $\langle\rho\rangle_{\phi}$ and $\langle\Omega\rangle_{\phi}/(2\pi)$. (c) Comparison between the dominant \nua{} $C_{11}$ half frequency, $f_{\mathrm{dom}}/2$ (orange squares and dashed line), and the time- and azimuthally averaged matter rotation frequency $\overline{\Omega}/(2\pi)$ evaluated at the radial location where $\langle\rho\rangle_{\phi}=10^{13}\,\mathrm{g\,cm^{-3}}$ (blue circles and solid line).
  Both frequency scales vary only weakly at late times but remain offset. (d) Inferred azimuthally averaged density regions, $\langle\rho\rangle_{\phi}$, associated with the $\nu_e$, $\bar{\nu}_e$, and $\nu_x$ signals in \texttt{SFHo\_M25960\_q100}. The density regions are obtained by matching $f_{\mathrm{dom}}/2$ to the temporally averaged $\overline{\Omega}/(2\pi)$--$\langle\rho\rangle_{\phi}$ relation. 
  All flavors cluster at $\langle\rho\rangle_{\phi}\simeq(5$--$7)\times10^{12}\, \mathrm{g\,cm^{-3}}$,
  suggesting an inner-disk origin for the late-time azimuthal modulation.}
\end{figure*}

To connect the time variability of the nonaxisymmetric dipole coefficient, \(C_{11}(t)\), with the dynamics of the remnant, we now inspect the equatorial angular–velocity field in the quasistationary phase (\(\gtrsim 20\) ms after merger). The rightmost plot in Fig.~\ref{fig:omega_in_polar_coord} shows \(\Omega/(2\pi)\) in polar coordinates for the same SFHo model, at the end of the simulation. The other plots present the time variation of the rotational profiles by computing the difference in $\Omega$ between selected times
and the final one.

At all times, the rotation is clearly differential. We recognize a slower central core, a broad annulus where the angular velocity peaks, and a decline toward the outer disk \citep{Kastaun_2015, Hanauske_2017, Cassing_2024}. The green isocontours (notably the \(\sim\!0.8\) kHz line) trace a ring that remains near a fixed radius over time.
With time, the degree of differential rotation tends to decrease, especially close to the peak.

To identify which densities correspond to the rotation rates, panels (a) and (b) of Fig.~\ref{fig:EF_density_radial_profile_signal_matter_freq_density_region} report the radial profiles of the angular-averaged equatorial rest-mass density $\langle\rho\rangle_{\phi}(r)$ and rotational frequency $\langle\Omega\rangle_{\phi}/(2\pi)$, together with their parametric relation. 
The maximum rotation rate occurs in the transition region between the remnant and the inner disk, at radii slightly larger than the azimuthally averaged equatorial \(10^{13}\,\mathrm{g\,cm^{-3}}\) isodensity contour.
We further compare this dynamical scale with the variability measured in the radial neutrino flux. Panel (c) of Fig.~\ref{fig:EF_density_radial_profile_signal_matter_freq_density_region} compares half of the dominant frequency extracted from 5-ms sliding-window FFTs of \(C_{11}(t)\) (orange squares) alongside the matter rotation frequency \(\overline{\Omega}/(2\pi)\) evaluated on the \(\langle\rho\rangle_{\phi}=10^{13}\,\mathrm{g\,cm^{-3}}\) isodensity surface (blue circles). After the early transient, the two curves track each other closely and remain within a narrow band for the rest of the evolution. 
This suggests that the persistent \(C_{11}\) oscillation could be set by the rotation of the near-surface inner-disk layers. 
To further explore this connection, panel (d) of Fig.~\ref{fig:EF_density_radial_profile_signal_matter_freq_density_region} shows for selected times and all neutrino flavors the average density at which the characteristic frequency of the $C_{11}$ coefficient divided by 2 equals the rotational frequency. 
The inferred densities cluster between \(5-7\times 10^{12}\,\mathrm{g\,cm^{-3}}\) with only mild secular drift and small flavor-to-flavor offsets. 
This indicates that the observed modulation does not originate across the last scattering neutrino surfaces. The latter are indeed characterized by a density of $\sim 10^{11} {\rm g~cm^{-3}}$ for all the neutrino species once the different opacities and mean energies are taken into account (see, e.g., \citep{Endrizzi_2020}). However, neutrino luminosities originate from an extended region across the last scattering surfaces and the contribution coming from a region of optical depth $\tau_{\nu}$ located $\Delta R$ below the surface can be estimated as
$ \delta L \sim e_{\nu}/t_{\rm diff} $ \citep{Ruffert1996,Rosswog2003},
where $e_{\nu}$ is the local neutrino energy density and $t_{\rm diff} \sim \Delta R ~\tau_{\nu} / c \sim \left( \Delta R \right)^2 ~\kappa_{\nu} / c$ is the diffusion timescale and $\kappa_{\nu}$ the total neutrino opacity. Around $ \tilde{\rho} \approx 5 \times 10^{12}{\rm g ~cm^{-3}}$, $\kappa \sim 1~{\rm km}^{-1}$ and
$$
t_{\rm diff} \sim 0.3 {\rm ms} \left( \frac{\Delta R}{10~{\rm km}} \right)^2 \left( \frac{\kappa_{\nu}}{1~{\rm km^{-1}}} \right) \lesssim \left. \left(\frac{\langle \Omega \rangle_{\phi}}{2 \pi} \right)^{-1} \right|_{\langle \rho \rangle_{\phi}= \tilde{\rho}} \, .
$$
Thus, the $C_{11}$ azimuthal modulation seems to track
a persistent anisotropy in the disk structure located in the semitransparent portion of the disk, where the diffusion timescale approaches the rotational timescale.


\section{Implications for Electron Fraction}
\label{sec:influence_electron_fraction}

The angular structure of the neutrino radial flux quantified in the previous sections has a direct impact on the composition of the material irradiated by the remnant.  In particular, charged-current absorption of electron neutrinos and antineutrinos,
$\nu_e+n\rightarrow p+e^-$ and $\bar{\nu}_e+p\rightarrow n+e^+$, drives the electron fraction toward an optically thin equilibrium value set by the relative neutrino luminosities and spectral energies.  We estimate this target value using Eq.~(\ref{eq:ye_asym}) \citep{Qian_1996}.  This estimate should not be interpreted as the final ejecta composition in all directions, since the actual $Y_e$ also depends on the local expansion time, density, temperature, and duration of neutrino exposure.  Rather, $Y_{e,\mathrm{eq}}$ provides a useful diagnostic of the composition toward which matter would evolve if neutrino absorption remained efficient for sufficiently long times.

Recent long-term M1 BNS merger simulations modeling outflows from long-lived neutron star remnants, see e.g. \citep{Bernuzzi_2025,Sneppen_2026,Jacobi_2026}, showed a significantly increased $Y_e$ in the ejecta, especially in the polar regions, reaching values even between 0.5 and 0.6. The amount of matter characterized by such high values of $Y_e$ has potential observational implications.
\citet{Sneppen_2026} investigate helium enrichment as a potential probe of the neutron-star remnant lifetime, while \citet{Jacobi_2026} find that high-$Y_e$ polar outflows from long-lived remnants can produce $^{56}\mathrm{Ni}$ and alter the subsequent radioactive heating and kilonova emission.
Comparing $Y_e$ with $Y_{e,\mathrm{eq}}$ helps distinguish the role of neutrino irradiation and assess the robustness of the results obtained in these simulations.

To connect this diagnostic with the angular structure of the radial flux, we compute $Y_{e,\mathrm{eq}}$ in 5-ms bins using the time-averaged electron-neutrino and electron-antineutrino energies on the extraction sphere. The luminosity ratio is obtained from the binned energy fluxes, while the spectral parameters entering Eq.~(\ref{eq:ye_asym}) are estimated from the corresponding energy-to-number flux ratios using the closure described in Sec.~\ref{sec:pa_methods_neutrino}. 
Regional averages quoted below are weighted by the detector surface element, with polar regions defined by $|\mathrm{\theta}|\geq 60^\circ$ and equatorial regions by $|\mathrm{\theta}|\leq 15^\circ$.  We verified that computing $Y_{e,\mathrm{eq}}$ from binned neutrino fields gives results consistent with averaging instantaneous $Y_{e,\mathrm{eq}}$ values over the same bins during the quasi-stationary phase; the largest differences occur during the first few milliseconds after merger, when the radial radiation flux is still rapidly evolving.

\begin{figure}
  \centering
  \includegraphics[width=1\linewidth]{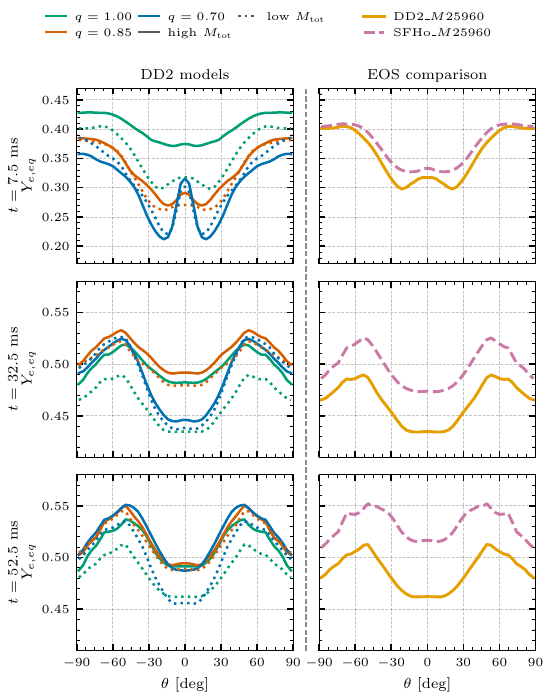}
  \caption{
  \label{fig:ye_eq_binned_theta_profiles}
  Polar-angle profiles of the neutrino-equilibrium electron fraction $Y_{e,\mathrm{eq}}$ computed from 5-ms retarded postmerger bins. The profiles are obtained from the binned $\nu_e$ and $\bar{\nu}_e$ fields using Eq.~(\ref{eq:ye_asym}); azimuthal averages are weighted by the detector surface element. The selected bins centered at $t=7.5$, $32.5$, and $52.5$ ms show the evolution from a pronounced early equatorial minimum to a milder late-time polar--equatorial contrast.
  }
\end{figure}

Figure~\ref{fig:ye_eq_binned_theta_profiles} shows the latitude dependence of $Y_{e,\mathrm{eq}}$ for representative 5-ms bins centered at $t=7.5$, $32.5$, and $52.5$ ms. The profiles inherit the large-scale geometry identified in Secs.~\ref{sec:angular_distribution} and \ref{sec:spherical_harm_decomp}: polar directions are exposed to a systematically more protonizing neutrino environment than equatorial directions.  At early times, the equatorial minimum is pronounced. In the 7.5-ms bin, the surface-element weighted polar-to-equatorial ratio
\(\langle Y_{e,\mathrm{eq}}\rangle_{\rm pole}/\langle Y_{e,\mathrm{eq}}\rangle_{\rm eq}\)
spans \(\simeq 1.15\)--\(1.41\) across the DD2 models. This early-time contrast reflects both the strong polar irradiation and the rapidly varying post-merger emission immediately after the hydrodynamical transient.

At later times, the entire angular distribution shifts toward larger equilibrium electron fractions. By \(t\simeq32.5\)--\(52.5\) ms, polar values are typically \(Y_{e,\mathrm{eq}}\simeq0.50\)--\(0.55\), while equatorial values rise to \(\simeq0.43\)--\(0.52\). The angular contrast therefore remains present but becomes milder: the surface-element weighted polar-to-equatorial ratio decreases to \(\langle Y_{e,\mathrm{eq}}\rangle_{\rm pole}/\langle Y_{e,\mathrm{eq}}\rangle_{\rm eq}\simeq1.03\)--\(1.18\). This behavior is important: although the energy-flux anisotropy is large, with polar-to-equatorial flux ratios of several, the corresponding anisotropy in $Y_{e,\mathrm{eq}}$ is moderated because the equilibrium value depends primarily on the relative $\nu_e$ and $\bar{\nu}_e$ capture efficiencies rather than on the absolute irradiation strength alone.  Thus, the quadrupolar radiation geometry strongly controls the neutrino exposure, but the composition target is set by the flavor-dependent balance of luminosities and mean energies.

The dependence on binary parameters follows the trends established earlier for the neutrino emission.  Unequal-mass DD2 systems generally display deeper equatorial minima and larger early-time polar--equatorial contrasts, consistent with stronger equatorial baryon loading and disk shadowing.  Increasing the total mass raises the overall equilibrium values, particularly at late times, reflecting the hotter and more compact remnants discussed in Sec.~\ref{sec:neutrino_luminosity_time_evolution}. 
The EoS comparison shows that BNS merger models using the softer SFHo EoS yield systematically larger \(Y_{e,\mathrm{eq}}\) at fixed low mass and equal mass ratio, especially during the late quasistationary phase. However, because these models also yield higher equatorial equilibrium values, their polar--equatorial contrast in \(Y_{e,\mathrm{eq}}\) need not be larger than in DD2, despite their stronger neutrino-flux anisotropy.

\begin{figure*}
  \centering
  \includegraphics[width=1\linewidth]{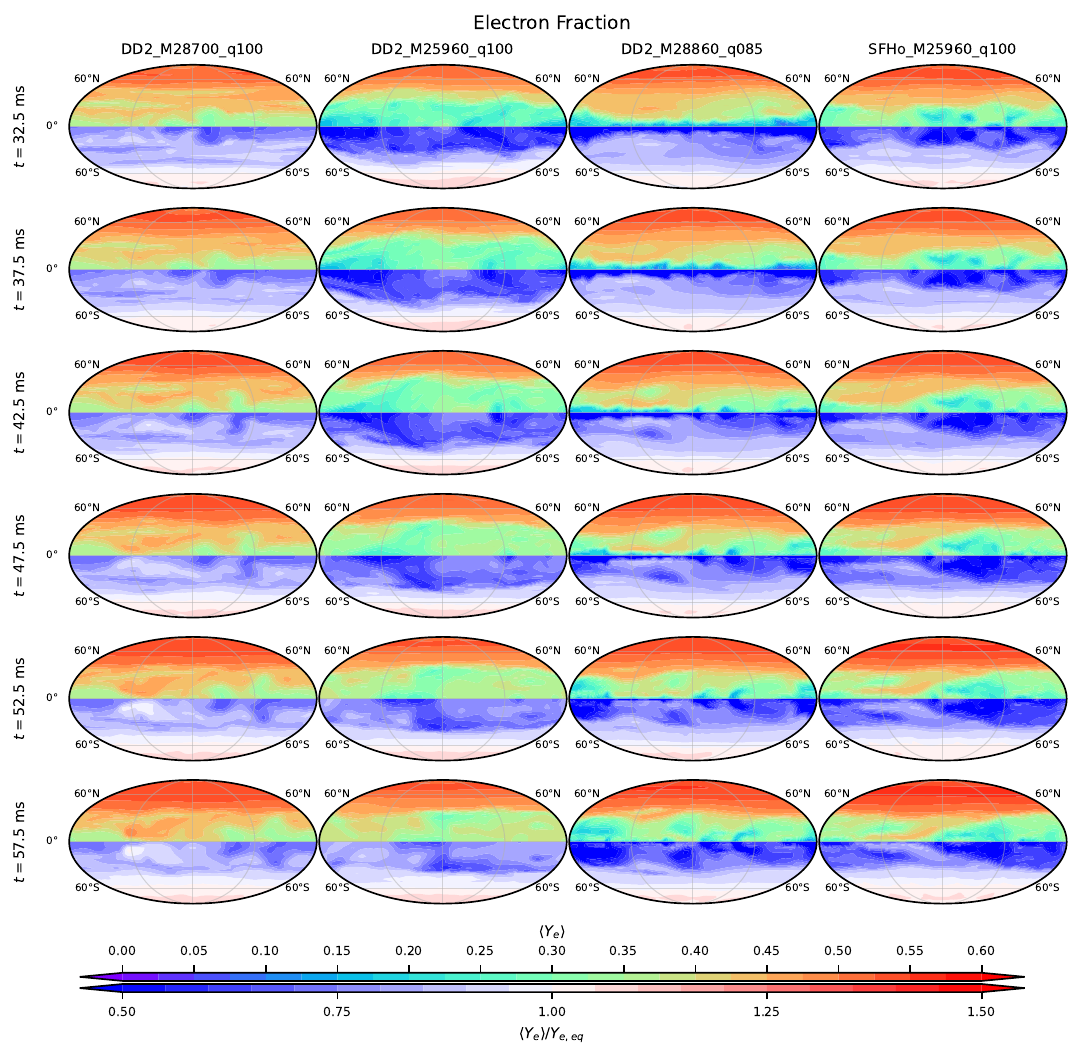}
  \caption{
  \label{fig:ye_eq_Ye_ratio_comparison}
  Binned angular comparison between the simulated electron fraction and the neutrino-equilibrium target for four representative models. Each row corresponds to a 5-ms retarded postmerger bin, and each column to a different binary configuration. In each Mollweide map, the northern hemisphere shows the binned electron fraction $\langle Y_e\rangle$, while the southern hemisphere shows the ratio $\langle Y_e\rangle/Y_{e,\mathrm{eq}}$. Values close to unity indicate material close to the neutrino-equilibrium target, whereas values below unity identify regions that remain less processed by neutrino absorption.
  }
\end{figure*}

Figure~\ref{fig:ye_eq_Ye_ratio_comparison} compares the simulated electron fraction with the neutrino-equilibrium target for four representative models over the late postmerger interval.  In each Mollweide panel, the northern hemisphere shows the binned electron fraction $\langle Y_e\rangle$, while the southern hemisphere shows the ratio $\langle Y_e\rangle/Y_{e,\mathrm{eq}}$. Polar regions generally have $\langle Y_e\rangle/Y_{e,\mathrm{eq}}\gtrsim1$, indicating that material exposed to the polar neutrino flux approaches or even marginally exceeds the equilibrium value.  In contrast, equatorial regions remain systematically below equilibrium, with typical ratios $\sim0.55$--$0.80$ depending on model and time.  The equatorial material is therefore not simply characterized by a lower equilibrium value; it is also less efficiently processed toward that equilibrium, consistent with reduced irradiation, larger optical depths, and shorter effective weak-interaction exposure before freeze-out.

These results complete the connection between the multipolar neutrino radiation and the composition of the outflow.  The monopole-plus-quadrupole structure identified in Sec.~\ref{sec:spherical_harm_decomp} produces a robust latitude-dependent weak-interaction environment: polar directions are driven toward higher $Y_e$, while equatorial directions remain more neutron rich.  This naturally supports the standard interpretation of viewing-angle-dependent kilonova emission, in which polar material is more favorable to lanthanide-poor, bluer emission, whereas equatorial material remains more lanthanide rich and contributes to redder components \citep{Metzger2010b,Kasen2017,Perego2017b,Radice2018}.  At the same time, the modest amplitude of the $Y_{e,\mathrm{eq}}$ contrast compared with the much larger flux contrast shows that composition cannot be inferred from the energy-flux anisotropy alone; the flavor and spectral structure of the radial radiation flux are essential.

The condition $Y_e \gtrsim Y_{e,{\rm eq}}$ in the polar region could have different explanations. It could indicate a small, yet non-negligible role of neutrino emission under optically thin conditions in setting the ejecta composition. Although this effect would be most visible in the polar funnel due to its approach toward irradiation-driven equilibrium, neutrino emission may exert a greater relative impact on the composition of the more neutron-rich and less irradiated equatorial ejecta. Alternatively, it could indicate that the angular structure of the radial flux changes slightly when moving from the remnant scale (where finite size effects are present and the irradiation is more effective) to infinity.

Finally, the small nonaxisymmetric variability associated with the $m=1$ mode, discussed in Sec.~\ref{sec:spherical_harm_decomp}, is subdominant for the time-averaged equilibrium composition considered here.  Its presence may introduce local azimuthal fluctuations, but the leading-order composition imprint is set by the axisymmetric polar--equatorial contrast.  In this sense, the late-time neutrino flux provides a compact and physically interpretable input for composition-aware kilonova modeling: a dominant monopole component, a robust quadrupolar correction controlling the viewing-angle dependence, and weaker rotational variability superposed on this large-scale geometry.

\section{Conclusions}
\label{sec:conclusions}

We analyzed the angle-integrated and angle-dependent neutrino emission from the simulated binary neutron-star mergers using the energy-integrated M1 scheme extracted on a sphere at $R\simeq295$ km. The results of Secs.~\ref{sec:neutrino_luminosity_time_evolution}--\ref{sec:influence_electron_fraction} can be summarized as follows.

\begin{itemize}

\item The angle-integrated luminosities rise to a merger peak of a few $10^{53}\,\mathrm{erg\,s^{-1}}$ and then decline during the simulated postmerger evolution. The luminosity ordering is $L_{\bar{\nu}_e}>L_{\nu_e}>L_{\nu_x}$, whereas the mean-energy ordering is $\langle E_{\nu_x}\rangle>\langle E_{\bar{\nu}_e}\rangle>\langle E_{\nu_e}\rangle$. In the displayed comparisons, the more massive equal-mass DD2 model has larger early luminosity peaks, and the low-mass equal-mass SFHo model has a harder spectrum, most clearly for $\nu_x$.

\item The angular maps show enhanced high-latitude fluxes and an equatorial flux suppression. The early postmerger maps contain substantial azimuthal structure; by $t\simeq15$--$60$ ms the large-scale latitude dependence is the dominant feature. The $\nu_x$ mean energy is larger than the electron-flavor mean energies throughout the displayed snapshots, while its angular contrast is weaker and evolves more gradually.

\item The polar-angle luminosity profiles of $\nu_e$ and $\bar{\nu}_e$ have minima near the equator and maxima toward the poles. In the DD2 sample, the polar-to-equatorial contrast increases as the mass ratio decreases: the $q=0.70$ models have larger ratios than the equal-mass models. At fixed $M=2.60\,M_\odot$ and $q=1$, the DD2 and SFHo ratios are similar at 7.5 ms and are larger for SFHo at 32.5 and 52.5 ms.

\item The late-time spherical-harmonic spectra are led by the $\ell=2$ component, while the $\ell=1$ contribution is smaller. For the fixed-low-mass comparison, SFHo has larger electron-flavor $P_2$ and $C_{20}$ values than DD2 at the three reported epochs. The $\nu_x$ energy flux also has a non-negligible $\ell=3$ axisymmetric contribution. The $C_{11}$ analysis identifies an $m=1$ modulation: after about 30 ms, the plotted half frequency of the DD2 models is $\simeq0.6$--$0.7$ kHz, whereas the SFHo model follows a different late-time evolution. For the SFHo model, matching the adopted half frequency to the rotation profile points to a portion of the inner disk characterized by rest mass densities of $(5$--$7)\times10^{12}\,\mathrm{g\,cm^{-3}}$. This kinematic association places the origin of the modulation in the regions below the last scattering neutrino surfaces where a significant fraction of the luminosity observed at infinity is generated.

\item The 5-ms binned $Y_{e,\mathrm{eq}}$ profiles are larger at high latitude than near the equator. Across the DD2 models, the surface-element weighted polar-to-equatorial ratio is $\simeq1.15$--$1.41$ in the 7.5-ms bin and decreases to $\simeq1.03$--$1.18$ at $t\simeq32.5$--$52.5$ ms. The binned $\langle Y_e\rangle/Y_{e,\mathrm{eq}}$ maps are generally close to unity at high latitude and remain below unity in equatorial regions.
\end{itemize}

Several limitations remain. The present analysis employs an energy-integrated (gray) M1 transport scheme. Energy-dependent treatments may modify the neutrino luminosities and mean energies as well as the disk and ejecta properties, e.g., \cite{Cheong2024}. The simulations also include a finite set of neutrino--matter reactions and do not account for all potentially relevant interaction channels and microphysical corrections. More complete rate treatments provide a path for future improvements, e.g., \cite{Chiesa2025,Rath2026,FoucartRath2026}. In addition, the simulations do not include self-consistent neutrino flavor evolution, whose possible impact on the merger dynamics and ejecta composition has begun to be explored, e.g., \cite{Qiu_2025a,Qiu_2025b}.

More broadly, the possible association between remnant rotational dynamics and neutrino emission variability highlights the importance of interpreting gravitational-wave, neutrino, and electromagnetic signals within a unified framework. The angular radiation structure quantified here provides a physically grounded basis for such multimessenger investigations of neutron star mergers.

Finally, we must acknowledge the inherent limitations of the M1 moment scheme employed in this study, which could impact the finer details of the angular distribution. The M1 approximation relies on an analytic closure relation (the Minerbo closure) that accurately captures both the optically thick diffusion limit and the bulk transition to free streaming. However, it is well known to struggle in optically thin regimes where radiation beams from different emission surfaces intersect (ray-crossing artifacts), e.g., \cite{Foucart2018,Foucart_2023}. In such regions, the closure can artificially shock or overly diffuse the radiation field, rather than allowing intersecting beams to pass through one another. Consequently, some fraction of the angular smoothing or minor higher-order multipole features in the optically thin limit could be influenced by the closure relation.
Although the remnant--disk geometry provides a physical basis for a polar--equatorial contrast, its quantitative amplitude, including the $P/E$ ratios, may be affected by artifact induced by the closure and the gray character of the transport. Angle- and energy-dependent transport calculations, including Monte Carlo comparisons for the same models, are needed to assess these systematic uncertainties.

\begin{acknowledgments}
We thank the anonymous referee for their careful reading and constructive comments, which have significantly improved the quality and clarity of the manuscript.
The current study is part of the PhD thesis of KAC. 
This study was supported by the Scientific and Technological Research Council of Turkey (T\"{U}B\.{I}TAK 117F188 and 119F077). KAC thanks T\"{U}B\.{I}TAK for his Fellowship (2214-A and 2211-C).
KAC also acknowledges the Department of Physics of the University of Trento for its hospitality during the development of this project. KY gratefully acknowledges the Master and Fellows of Churchill College, University of Cambridge, for the award of an Overseas Research Fellowship.
FMG and AP are supported by the European Union under NextGenerationEU, PRIN
2022 Project No. 2022KX2Z3B. FMG and AP also acknowledge the EuroHPC Joint Undertaking for awarding this project access to the EuroHPC supercomputer
LUMI, hosted by CSC (Finland) and the LUMI consortium through a EuroHPC Extreme Scale Access call (EHPC-EXT-2022E01-046).
\end{acknowledgments}

\section*{Data Availability}
There are no publicly available research data or software supporting this manuscript. Requests for further information or data should be sent to the authors.

\appendix

\section{NEUTRINO EMISSION FROM LONG-LIVED AND PROMPT-COLLAPSE SFHO REMNANTS}
\label{app:sfho_remnant_outcome}

\begin{figure*}
  \centering
  \includegraphics[width=1\linewidth]{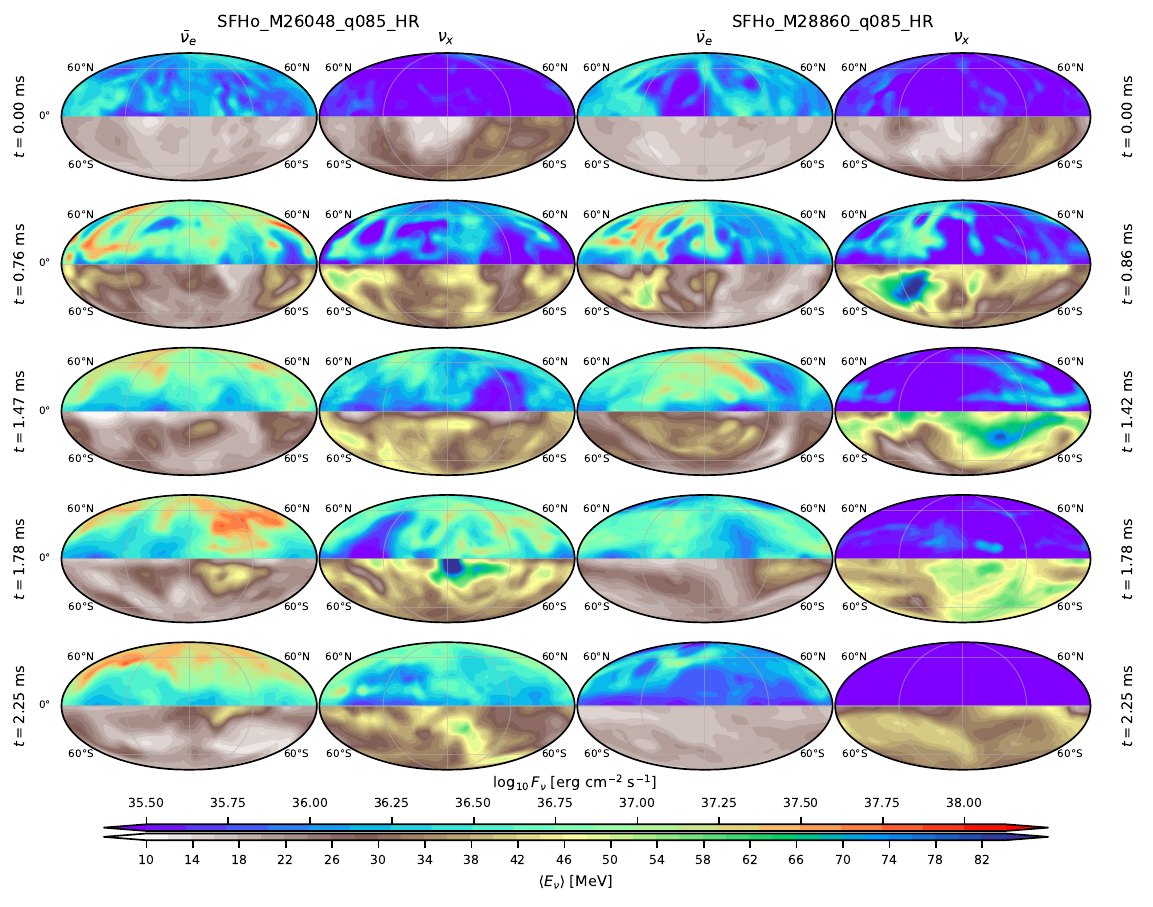}
  \caption{
    \label{fig:angular_dist_nu_SFHo_M26048_q085_HR_vs_SFHo_M28860_q085_HR}
    Same angular representation and color scales as in
    Fig.~\ref{fig:angular_dist_nu_DD2_M2870_q100}, but for two SFHo binaries
    with $q=0.85$ and different postmerger outcomes. The first two columns show
    $\bar{\nu}_e$ and $\nu_x$, respectively, for \model{SFHo\_M26048\_q085\_HR},
    which forms a long-lived massive neutron-star remnant. The third and fourth
    columns show the same species for \model{SFHo\_M28860\_q085\_HR}, which collapses
    to a black hole at $t\simeq0.9$~ms. Rows show snapshots near $t=0$,
    $0.8$, $1.4$--$1.5$, $1.78$, and $2.25$~ms; the exact times for the two
    models are indicated on the left and right sides of the figure. The northern
    displayed hemisphere gives $\log_{10}F_r$ in
    $\mathrm{erg\,cm^{-2}\,s^{-1}}$, while the southern displayed hemisphere
    gives the mean energy in MeV.
    }
\end{figure*}

This Appendix provides an illustrative comparison of the neutrino-emission
geometry in two SFHo binaries with the same mass ratio, $q=0.85$, but different total masses and postmerger outcomes (Fig.~\ref{fig:angular_dist_nu_SFHo_M26048_q085_HR_vs_SFHo_M28860_q085_HR}). The lower-mass model, \model{SFHo\_M26048\_q085\_HR}, forms a long-lived massive neutron-star remnant,
whereas \model{SFHo\_M28860\_q085\_HR} undergoes collapse to a black hole at
$t\simeq0.9$~ms. Because the two models also differ in total mass, this
comparison is intended to illustrate the contrast between persistent and
rapidly quenched neutrino emission rather than to isolate remnant fate as a
single controlled parameter.

Before collapse, both configurations show an anisotropic radial radiation flux with
enhanced high-latitude flux and reduced equatorial emission. In the long-lived model, this morphology persists throughout the displayed interval:
the initially patchy $\bar{\nu}_e$ and $\nu_x$ flux distributions evolve
toward a smoother, predominantly latitude-dependent structure while retaining
a clear polar--equatorial contrast. The $\nu_x$ channel remains harder than
the $\bar{\nu}_e$ channel, although its mean-energy distribution retains
substantial time-dependent angular structure.
The prompt-collapse configuration displays a qualitatively similar geometry
before black-hole formation, but its flux decreases rapidly after $t\simeq0.9$~ms. At $t\simeq1.4$--$2.25$~ms, both the $\bar{\nu}_e$ and,
especially, the $\nu_x$ fluxes are much weaker than in the long-lived model,
and the coherent polar enhancement is strongly reduced. Local $\nu_x$ mean
energies can remain high or fluctuate after collapse; because the associated
flux is already small, these values do not imply sustained neutrino irradiation.

This comparison shows that the early polar--equatorial geometry is established
before the two remnants diverge dynamically, whereas its temporal persistence
depends strongly on whether a massive neutron-star remnant survives. A
long-lived remnant can therefore maintain an anisotropic neutrino source for
continued weak processing of surrounding material, while prompt collapse
rapidly quenches this irradiation.


\bibliography{refs}

\end{document}